\documentclass[aps,reprint,prd,amsmath,amssymb,nofootinbib,showpacs]{revtex4-1}
\usepackage[english]{babel}
\usepackage{graphicx}
\usepackage{bm}
\usepackage{epsf}
\usepackage{dcolumn}

\usepackage{color}
\usepackage[normalem]{ulem}  

\newcommand{\new}[1]{#1}

\begin{document}

\title{Transport coefficients of nucleon neutron star cores for various nuclear interactions within the Brueckner-Hartree-Fock approach}
\author{P.~S. Shternin}\email{pshternin@gmail.com}
\affiliation{Ioffe Insitute, 26 Politekhnicheskaya st., St.\ Petersburg, 194021, Russia}
\author{M.~Baldo}
\affiliation{INFN Sezione \ di Catania, Via Santa Sofia 64, 95123 Catania, Italy}

\date{\today}


\begin{abstract}
We consider the thermal conductivity, shear viscosity, and momentum relaxation rates in the nucleon cores of the neutron stars. We study how the choice of the nuclear interaction and the model for three-body forces may affect these transport coefficients calculated within the Brueckner-Hartree-Fock many-body nuclear theory. We find that at relatively large densities the model dependence of the results is substantial. In addition we provide the analytical approximations which allow to incorporate our results in practical simulations.
\end{abstract}
\maketitle

\section{Introduction}\label{sec:intro}

Studies of matter at and beyond the nuclear density ($n_0\approx 0.16$~fm$^{-3}$) attract  constant interest as a way to test fundamental physical theories such as theories of strong interactions. 
One of a few  ways to study matter at supranuclear densities comes from the neutron star astrophysics.  Neutron stars (NSs), having masses of the order of the Solar mass $M_\odot$ and radii of the order of tens of kilometers, are largely composed of the superdense matter in their liquid cores, although the exact composition and equation of state (EOS) of this matter are unknown \cite{HPY2007Book}. In particular, the properties of the superdense matter inside NS cores affect the flow of various non-equilibrium processes that have observational consequences. 
Confronting the results of modeling of such processes with the astrophysical observations potentially allows one to infer the underlying physical properties deep in NS interiors. The important microphysical input to such a modelling are the transport properties of the NS matter. 

Transport properties of NS interiors were studied intensively in the last decades, see Ref.~\cite{Schmitt2018} for a review. Most of the results were obtained for a simplest nucleon composition of NS cores, where the matter constituents are neutrons (n), protons (p), electrons (e), and muons ($\mu$). The matter inside NSs is thought to be at, or close to the equilibrium with respect to the weak reactions, the so-called beta-stable matter. In this case the proton (and lepton) fraction $x_p$ is small, $x_p\lesssim 0.2-0.3$ \cite{HPY2007Book}. Depending on the transport problem considered, the largest  contribution to transport coefficients comes \new{from} neutrons or leptons, with the protons giving the least important contribution. 
The first detailed studies of the diffusive transport coefficients of NS cores were performed by \citet{FlowersItoh1979ApJ}. Their consideration of nucleon transport coefficients was based on the free-space scattering cross-sections. However, since nucleons inside NS cores form dense non-ideal strongly interacting Fermi liquid, the appropriate nuclear many-body theory needs to be incorporated. The approaches to this problem included the in-medium perturbation theory based on the Brueckner $G$-matrix \cite{Benhar2010PhRvC, Zhang2010PhRvC,Shternin2013PhRvC,Shternin2017JPhCS} or thermodynamical $T$-matrix \cite{Sedrakian1994PhLB}, variational approach within the correlated basis function formalism \cite{BenharValli2007PhRvL, Benhar2010PhRvC, CarboneBenhar2011JPhCS}, and  the medium-modified one-pion exchange model \cite{Blaschke2013PhRvC, Kolomeitsev2015PhRvC} in the framework of Landau-Migdal Fermi-liquid theory \cite{Migdal1990PhR}, see Ref.~\cite{Schmitt2018} for the more detailed review.

In the present study we 
continue our previous work \cite{Shternin2013PhRvC} and investigate the transport coefficients of the NS cores within the Brueckner-Hartree-Fock (BHF) framework. We limit ourselves to the simplest npe$\mu$ composition of the NS cores and focus on the nucleon contribution to the transport. For the up-to-date results on the lepton contribution, see Refs.~\cite{Schmitt2018,ShterninYakovlev2007, ShterninYakovlev2008, Shternin2008JETP, Shternin2018PRD}.
Previously \cite{Shternin2013PhRvC} 
we obtained the nucleon thermal conductivity $\kappa$ and shear viscosity $\eta$ within this model in the beta-stable matter of NS core for one particular nuclear interaction, namely the Argonne v18 \new{two-body} potential supplemented with the Urbana IX  three-body force. 
Here we explore the dependence of our results on the choice of the nuclear interaction following Ref.~\cite{Baldo2014PhRvC} where similar analysis was performed for the nucleon effective masses. In addition, we also consider the momentum relaxation rate in the neutron-proton collisions, an important transport coefficient in studies of the magnetic field evolution in NS cores \cite{Iakovlev1991Ap&SS, Yakovlev1991Ap&SS, Goldreich1992ApJ, Schmitt2018}. 

Preliminary results of the present studies were reported in Ref.~\cite{Shternin2017JPhCS}.
We find, that the dependence of the transport coefficients on the model of the interaction can be substantial at high baryon densities. The main difference seems to come from the model for three-body forces. While at low densities, $n_B\sim n_0$,
all calculations agree, at higher densities values of thermal conductivity, shear viscosity, and momentum relaxation rate can differ by an order of magnitude depending on the selected NN interaction.

The paper is organized as follows. In Section~\ref{sec:formalism} we outline the formalism used to calculate the kinetic coefficients and introduce the effective mean free paths of the transport theory. In Section~\ref{sec:bhf} we shortly review the BHF approach and we present our results in Section~\ref{sec:discuss}. \new{Specifically, in} Section~\ref{sec:mfp} we analyse the dependence of the effective mean free paths on the selection of the nuclear interaction. In Section~\ref{sec:exact} we discuss the simplifications that can be used in calculations of the shear viscosity and thermal conductivity.  In Section~\ref{sec:partial} we try to isolate main partial waves of the nucleon interaction that are responsible for the difference in the results. Finally in Section~\ref{sec:pract} we provide the practical expressions for calculating the transport coefficients in our model. We conclude in Section~\ref{sec:conclude}.

In what follows,  we set $\hbar=k_B=c=1$, except for the practical expressions in Section~\ref{sec:pract}.
Effects of nucleon pairing are outside the scope of the present paper and are not included.

\section{Formalism}\label{sec:formalism}
The calculation of the transport coefficients in NS cores is based on the transport theory of multi-component Fermi liquids \cite{BaymPethick,Anderson1987,FlowersItoh1979ApJ}. We consider here the thermal conductivity coefficient $\kappa_i$, shear viscosity coefficient $\eta_i$, and momentum relaxation rate $J_{ij}$, where indices $i$ and $j$ numerate particle species. It is customary to express these transport coefficients through the effective mean free paths of the quasiparticles as
\begin{eqnarray}
  \kappa_i &=& \frac{\pi^2}{3} T \frac{n_i}{p_{Fi}} \lambda_i^\kappa, \label{eq:kappa}\\
  \eta_i &=&  \frac{1}{5} n_i p_{Fi} \lambda_i^\eta, \label{eq:eta}\\
  J_{ij} &=& n_i p_{Fj} \left(\lambda_{ij}^{D}\right)^{-1},\label{eq:J}
\end{eqnarray}
where $T$ is a temperature, $n_i$ is a number density of the particle species i, $p_{Fi}=(3\pi^2 n_i)^{1/3}$ is the corresponding Fermi momentum. In Equations~(\ref{eq:kappa})--(\ref{eq:eta}),  the quantities $\lambda^\kappa_i$ and $\lambda^\eta_i$ are effective mean free paths of the particle species i for the thermal conductivity and shear viscosity problems, respectively, while $\lambda^D_{ij}$ introduced in Equation~(\ref{eq:J}) is an effective mean free path for the momentum relaxation in the collisions  between particle species $i$ and $j$.
 Momentum relaxation rates (\ref{eq:J}) describe friction between the mixture components. They are related to the traditional diffusion coefficients as $D_{ij}=n_i \mu_i (\partial \log \mu_i/\partial \log n_i) J_{ij}^{-1}$\new{, where $\mu_i$ is a chemical potential of the particle species i}.
The effective mean free paths in Equations~(\ref{eq:kappa})--(\ref{eq:J}), in general, are not the same and need to be determined from microscopic calculations for the corresponding transport problem. Frequently, one expresses the transport coefficients through the effective relaxation times $\tau_i$ instead of $\lambda_i$ \cite{Schmitt2018}. The relation between these quantities is simply $\lambda_i=v_{Fi}\tau_i$, where $v_{Fi}$ is the Fermi velocity of the particle species i.

In order to find effective mean free paths in the degenerate matter of NS cores it is enough to employ the simplest variational solution of the transport equation, which \new{for thermal conductivity and shear viscosity problems} reduces to a solution of a system of linear equations, see Ref.~\cite{Schmitt2018} for details. This system reads
\begin{equation}\label{eq:vareq}
1=\sum\limits_j \left(n_j\sigma_{ij} \lambda_i + n_j \sigma_{ij}' \lambda_j\right),
\end{equation}
where the summation is carried over all components with which the given species $i$ collide, including $j=i$; $\sigma_{ij}$ and $\sigma_{ij}'$ are the transport cross-sections for these collisions to be defined below, and we omitted for a moment the upper indices at $\lambda$'s for brevity. The primed cross-sections $\sigma_{ij}'$ describe the mutual influence of the non-equilibrium distributions of the different particle species.
It is instructive to introduce a partial mean free path for the collisions between the particle species $i$ and $j$
\begin{equation}\label{eq:lambda_partial}
\lambda_{ij}\equiv \left(n_j \sigma_{ij}+\delta_{ij} n_j\sigma_{ij}'\right)^{-1},
\end{equation}
where $\delta_{ij}$ is the Kroenecker delta-symbol. The smaller the partial mean free path is, the more important is the corresponding scattering channel. We also define here the primed partial mean free paths that correspond to the non-diagonal terms in the system~(\ref{eq:vareq}):
\begin{equation}\label{eq:lambda_partial_prime}
    \lambda_{ij}'\equiv \left(n_j \sigma_{ij}'\right)^{-1}.
\end{equation}

In principle, all possible pair collisions in the mixture should be included in Equation~(\ref{eq:vareq}). However, it turns out (e.g., \cite{FlowersItoh1979ApJ,Schmitt2018}) that in the npe$\mu$ NS cores,  the system of equations (\ref{eq:vareq}) decouples in two subsystems, one of which corresponds to the nuclear sector and other to the electromagnetic one. These sectors can be considered separately. Moreover, only the neutrons as most abundant particles, dominate the nuclear contribution to the transport coefficients.  We will discuss this statement in more details in Section~\ref{sec:exact}.
In the present study we focus on the nucleon sector. We closely follow Refs.~\cite{Shternin2013PhRvC,Shternin2017JPhCS} and omit the details. 

In order to find the transport cross-sections that appear in Equation~(\ref{eq:vareq}), one needs to multiply  the quasiparticle collision probabilities with certain  angular factors depending on the transport problem in question and average the results over the allowed phasespace, see, e.g.,  Ref.~\cite{Shternin2013PhRvC}. The collision probability depends on the incoming pair particle state $|ij\rangle$,  and the final sate of the particles after the collisions $\langle i'j'|$. 
Since the particles are degenerate, only the excitations close to the Fermi surface contribute to the transport. Therefore, the magnitudes of all four particle momenta participating in the collision ($\mathbf{p}_i$, $\mathbf{p}_j$, $\mathbf{p}_{i'}$, $\mathbf{p}_{j'}$) are fixed to the respective Fermi momenta. Taking into account the conservation of the total momentum $\mathbf{P}\equiv\mathbf{p}_{i}+\mathbf{p}_j=\mathbf{p}_{i'}+\mathbf{p}_{j'}$, only two angular variables determine the relative position of the scattering quasiparticle momenta in space. For the nuclear scattering, it is convenient to use the absolute value of the total momentum, $P$, and the value of the transferred momentum, $q$, where $\mathbf{q}=\mathbf{p}_{i'}-\mathbf{p}_i=\mathbf{p}_{j}-\mathbf{p}_{j'}$. The latter is connected to the center of mass (c.m.) scattering angle $\theta_{\mathrm{c.m.}}$ as 
\begin{equation}\label{eq:theta_cm}
  \cos \theta_{\mathrm{c.m.}}=1-\frac{q^2}{2p^2},
\end{equation}
where $p$ is the absolute value of the colliding pair relative 
momentum $\mathbf{p}\equiv(\mathbf{p}_{j}-\mathbf{p}_{i})/2$. 
Since all quasiparticles are placed on the Fermi surface,  the relation
$4p^2+P^2=2(p_{Fi}^2+p_{Fj}^2)$ holds. Notice, that 
sometimes the other set of variables (e.g. the traditional Abikosov-Khalatnikov angles \cite{BaymPethick}) can be more convenient depending on the investigated problem. For instance, for the electromagnetic collisions it is more convenient to use the angle between $(\mathbf{p}_{i}\mathbf{p}_{i'})$ and $(\mathbf{p}_{j}\mathbf{p}_{j'})$ planes instead of $P$ (see, e.g., Ref.~\cite{Shternin2018PRD}).

Let $\langle
i'j'|\hat{G}|ij\rangle$ be the scattering transition matrix element, where $\hat{G}$ is the scattering  operator. We define 
\begin{equation}\label{eq:Qci}
{\cal Q}_{ij}(P,q)=\frac{1}{4(1+\delta_{ij})} \sum\limits_{\mathrm{spins}}|\langle
i'j'|\hat{G}|ij\rangle|^2,
\end{equation}
where the summation is carried over the initial and final spin states. The factor $(1+\delta_{ij})^{-1}$ is included to avoid double counting of the same collision events in the final expressions\footnote{Notice, that this differs from the definition in Refs.~\cite{Baiko2001AA,Shternin2013PhRvC}, where this factor is included at the later stage}. We also introduce the phase-space angular brackets for an arbitrary quantity \new{$F(P,q)$} as
\begin{equation}\label{eq:angav}
  \left\langle F(P,q) \right\rangle =
  \int\limits_{|p_{Fi}-p_{ Fj}|}^{p_{ Fi}+p_{Fj}} {\rm d} P
  \int\limits_0^{q_m(P)}\, \mathrm{d} q   \frac{F(P,q)}{\sqrt{q_m^2-q^2}},
\end{equation}
where $q_m(P)$ is the maximal possible transferred momentum for a given value of $P$. It can be expressed as \cite{Baiko2001AA}
\begin{equation}\label{eq:qm}
  q^2_m(P)=\frac{4p_{Fi}^2 p_{Fj}^2 - (p_{ Fi}^2+p_{ Fj}^2-P^2)^2}{P^2}.
\end{equation}

Using the definitions (\ref{eq:Qci})--(\ref{eq:angav}), the transport cross-sections for the thermal conductivity problem can be written as \cite{Shternin2017JPhCS}
\begin{eqnarray}
\sigma_{ij}^\kappa &=&\frac{3 m_i^{*2} m_j^{*2} T^2}{10 p_{Fi}^4p_{Fj}^3} 
\left\langle{\cal Q}_{ij}(4p_{Fi}^2+q^2)\right\rangle \label{eq:sigmakappa},\\
{\sigma_{ij}'}^{\kappa}&=& \frac{3 m_i^{*2}m_j^{*2} T^2}{10 p_{Fi}^3 p_{Fj}^4 }
\left\langle {\cal Q}_{ij}\left(2p_{Fi}^2+2p_{Fj}^2-2P^2-q^2\right)\right\rangle, \nonumber\\
\ \label{eq:sigmakappaprime}
\end{eqnarray}
where $m^*_{i,j}$ are the particle effective masses at the Fermi surface, so that $v_{Fi}=p_{Fi}/m_i^*$. Similar expressions for the shear viscosity problem, albeit different angular factors, read
\begin{eqnarray}
\sigma_{ij}^\eta &=&\frac{3 m_i^{*2} m_j^{*2} T^2}{8 p_{F}^6p_{Fj}^3} 
\left\langle{\cal Q}_{ij} q^2(4p_{Fi}^2-q^2)\right\rangle, \label{eq:sigmaeta}\\
{\sigma_{ij}'}^{\eta}&=& \frac{3 m_i^{*2}m_j^{*2} T^2}{8 p_{Fi}^5 p_{Fj}^4 }
\left\langle {\cal Q}_{ij}q^2\left(2p_{Fi}^2+2p_{Fj}^2-2P^2-q^2\right)\right\rangle.\nonumber\\
\ \label{eq:sigmaetaprime}
\end{eqnarray}
The effective mean free path 
for the momentum relaxation rate (\ref{eq:J}) can also be expressed via the corresponding transport cross-section 
\begin{equation}
\left(\lambda_{ij}^D n_j\right)^{-1}=\sigma_{ij}^D=\frac{m_i^{*2} m_j^{*2}T^2
}{2 p_{Fi}^4 p_{Fj}^3} 
\left\langle{\cal Q}_{ij} q^2\right\rangle.\label{eq:sigmaD}
\end{equation}
Notice that Equations~(\ref{eq:J}) and (\ref{eq:sigmaD}) imply that the momentum relaxation rates are symmetric, $J_{ij}=J_{ji}$. 
If the forward scattering, $q\to 0$,  dominates, 
\new{Equations~(\ref{eq:sigmaeta}) and (\ref{eq:sigmaD}) lead to the relation}
$\sigma_{ij}^\eta=3\sigma_{ij}^D$  \cite{Heiselberg:1993cr}. 
In the traditional Fermi-liquids, the squared transition amplitude (\ref{eq:Qci}) does not depend on the  energy transferred in the collisions. In this case the transport cross-sections in Equations~(\ref{eq:sigmakappa})--(\ref{eq:sigmaD}) obey the standard $\propto T^2$  temperature dependence. As a consequence, the effective mean free paths are inversely proportional to the temperature squared. This is the case for the nucleon collisions which are the main focus of the present study. For the long-range electromagnetic collisions in relativistic matter, this is no longer the case (see. e.g., Ref.~\cite{Schmitt2018} for the review).

Thus, to calculate the nucleon transport cross-sections one needs the squared matrix elements ${\cal Q}_{ij}$ and the effective masses $m^*_{i,j}$ which 
should  be provided by the nuclear many-body theory. Nuclear potentials and hence the scattering amplitudes are conveniently given in a partial wave basis for the interacting pair states in the c.m. frame, $|P,p; J\ell S M\rangle$, where $S$ is the pair total spin, $\ell$ is the pair orbital momentum, $J$ is its total angular momentum, and $M$ is the total angular momentum projection. Then the quantity ${\cal Q}_{ij}$ can be expanded in the series in Legendre polynomials
${\cal P}_L \left(\cos \theta_{\rm cm}\right)$:
\begin{equation}\label{eq:Q_legendre}
  {\cal Q}_{ij}(q,P)=\frac{1}{1+\delta_{ij}}\sum_{L} {\cal Q}^{(L)}_{ij}(P) {\cal
  P}_L\left(\cos \theta_{\rm cm}\right),
\end{equation}
where the coefficients of expansion are related to the matrix elements of the transition amplitude in the partial wave basis as  \cite{Shternin2013PhRvC} 
\clearpage
\begin{widetext}
\begin{eqnarray}
  {\cal Q}_{ij}^{(L)}(P)&=& \frac{1}{16 \pi^2}\sum i^{\ell'-\ell+\bar{\ell}-\bar{\ell}'}
  \Pi_{\ell\ell'}\Pi_{\bar{\ell}\bar{\ell}'}\Pi^2_{J\bar{J}}
  C^{L'0}_{\ell' 0 \bar{\ell}'0} C^{L0}_{\ell0\bar{\ell}0}
  \left\{\begin{array}{ccc}
    \bar{\ell} & S & \bar{J}\\
    J & L & \ell
  \end{array}\right\}
 \left\{\begin{array}{ccc}
    \bar{\ell}' & S & \bar{J}\\
    J & L & \ell'
  \end{array}\right\}\nonumber\\
  &&\times
  \left(1+\delta_{ij}(-1)^{S+\ell}\right)\left(1+\delta_{ij}(-1)^{S+\bar{\ell}}\right)
  G^{JS}_{\ell\ell'}(P,p,p) \left(G^{\bar{J} S}
  _{\bar{\ell}\bar{\ell}'}(P,p,p)\right)^*\label{eq:Q_L}.
\end{eqnarray}
\end{widetext}
Here  $C^{L0}_{\ell0\bar{\ell}0}$ is the Clebsch-Gordan coefficient, terms in curly brackets are 6$j$-symbols of the quantum angular momentum theory \cite{Varshalovich},
$\Pi_{fg}\equiv\sqrt{(2f+1)(2g+1)}$,  $G^{JS}_{\ell\ell'}(P,p,p)$ is the matrix element of the operator $\hat{G}$.
In Equation~(\ref{eq:Q_L}), the total angular momentum $J$, total nucleon pair spin $S$, and total momentum $P$ are conserved, and summation is carried over
all angular momenta and spin variables, except $L$. The collision type index, nn, np, or pp,
of the $G$-matrix is omitted for brevity. 
Terms in brackets containing $\delta_{ij}$  account for the contribution of exchange terms in case when the collisions between the same species are considered. Using the expansion (\ref{eq:Q_legendre}), the integration over $q$ in Equations~(\ref{eq:sigmakappa})--(\ref{eq:sigmaD}) can be performed analytically (see Appendix A in Ref.~\cite{Shternin2013PhRvC}), and only a single integration over $P$, that needs to be performed numerically, remains.

\section{Brueckner-Hartree-Fock approach}\label{sec:bhf}
The nuclear many-body theory used in this study is the  non-relativistic Brueckner-Hartree-Fock (BHF) theory \cite{Baldo1999Book}. In choice of this model we closely follow Refs.~\cite{Shternin2013PhRvC,Baldo2014PhRvC,Shternin2017JPhCS}.
In this approach, the in-medium scattering matrix, or $G$-matrix, is found from the solution of the Bethe-Goldstone equation 
\begin{equation}\label{eq:BBG}
G^\alpha[n_B;z]=V^\alpha+\sum\limits_{k_a,k_b} V^\alpha \frac{|k_a k_b\rangle Q^\alpha\langle k_a k_b|}{z-e_a(k_a)-e_b(k_b)} G^\alpha[n_B;z],
\end{equation}
where the index $\alpha\equiv ab = $ nn, np, or pp, specifies the scattering species,  $V^\alpha$ is the bare nucleon-nucleon (NN) interaction, $z$ is the starting energy, $Q^\alpha$ is the Pauli operator. Single-particle energy of the species $a$, $e_a(k)$, in Equation~(\ref{eq:BBG}) is 
\begin{eqnarray}
e_a(k_a)&=&\frac{k_a^2}{2 m_N} \nonumber\\
&+& \mathrm{Re}\sum\limits_{b,\, k_b\leq p_{\mathrm{F}b}} \langle k_ak_b|G^\alpha[n_B; e_a(k_a)+e_b(k_b)]|k_ak_b\rangle_A,\nonumber\\
\label{eq:SPP}
\end{eqnarray}
where \new{$m_N$ is the bare nucleon mass and the} subscript $A$ means antisymmetrization of the wavefunction. Since the single-particle energy depends on the  $G$-matrix, Bethe-Goldstone equation needs to be solved self-consistently in the iterative manner. 
In Equations~(\ref{eq:BBG})--(\ref{eq:SPP}) the so-called continuous choice of the single-particle potential is adopted \cite{Baldo1999Book}. The total binding energy per nucleon is then
\begin{equation}\label{eq:E/A}
    \frac{B}{A}=
    \frac{1}{2 n_B}\sum\limits_{ab} \sum\limits_{
    \begin{array}{l} \scriptscriptstyle k_a\leq p_{\mathrm{F}a}\\
    \scriptscriptstyle k_b\leq p_{\mathrm{F}b} \end{array}
    } 
    \left<k_a k_b|G^{\alpha}[n_B; e_a(k_a)+e_b(k_b)]|k_ak_b\right>_A,
\end{equation}
To get the total energy, the free kinetic energy part has to be added.
In the partial wave basis, the expression for the binding energy reads
\begin{eqnarray}
\frac{B}{A} &=&\, \frac{1}{4\pi^2 n_B} \, \sum\limits_{\alpha,\,\ell JS} (2J+1) \int \mathrm{d}\ \cos\theta\, 
\int_0^{p_{Fa}} k_a^2 \mathrm{d}k_a \nonumber\\
&& \int_0^{p_{Fb}}  k_b^2 \mathrm{d}k_b\  G^{\alpha,JS}_{\ell\ell}(P,p,p;e_a(k_a)+e_b(k_b)),
\label{eq:energy_part}
\end{eqnarray}
where $P=2^{-1}|\mathbf{k}_a+\mathbf{k}_b|$, $p=|\mathbf{k}_a-\mathbf{k}_b|$ and $\theta$ is the angle between $\mathbf{k}_a$ and $\mathbf{k}_b$. The summation over $\alpha$ here includes $\alpha=\text{nn},\, \text{np},\,\text{pn},\,\text{and } \text{pp}$. We do not show explicitly the isospin index $I$ in Equation~(\ref{eq:energy_part}). It is fixed by the condition $\ell+S+T$ being odd. For the nn and pp scattering only $I=1$ channels contribute to the sum in Equation~(\ref{eq:energy_part}), while $I=0,\,1$ contribute to the np $G$-matrix.

\begin{figure*}[t]
\begin{minipage}{0.48\textwidth}
\includegraphics[width=\textwidth]{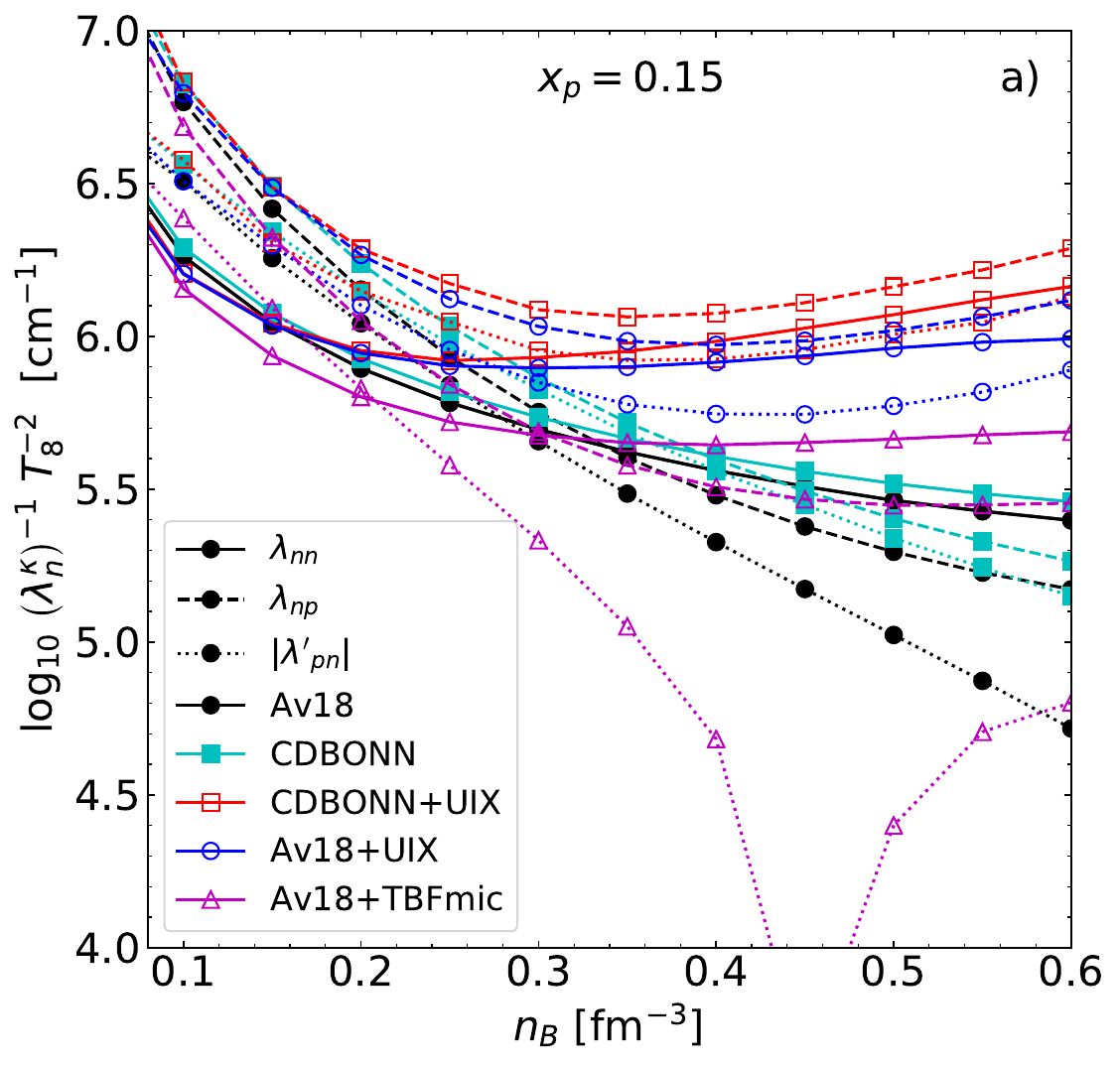}
\end{minipage}
\hspace{0.02\textwidth}
\begin{minipage}{0.48\textwidth}
\includegraphics[width=\textwidth]{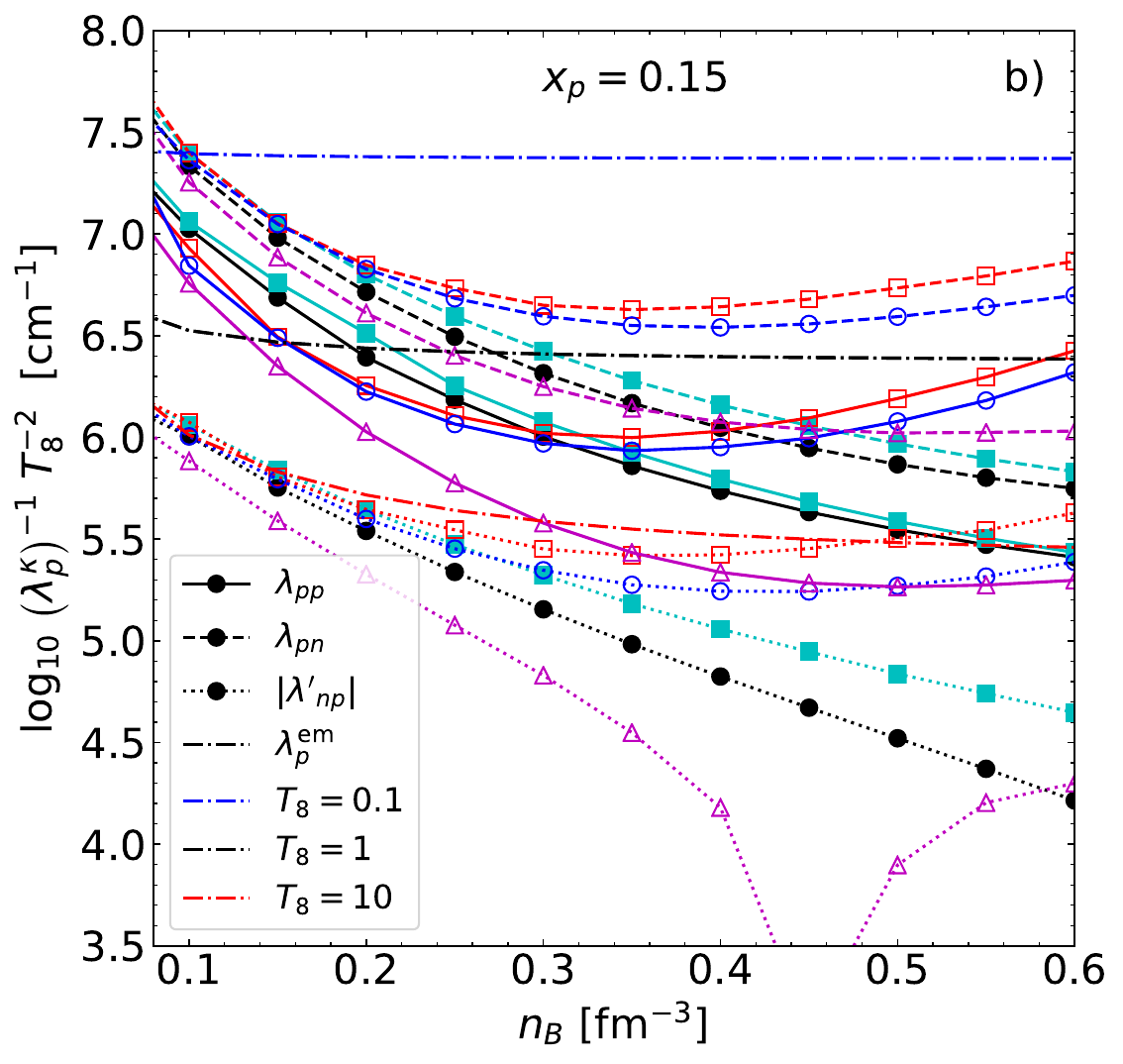}
\end{minipage}
\caption{Effective inverse mean free paths for the thermal conductivity problem as functions of density for the proton fraction $x_p=0.15$. The coefficients that couple to $\lambda^\kappa_n$ in Equation~(\ref{eq:vareq}) are shown in the  panel (a), while coefficients which couple to $\lambda^\kappa_p$ in Equation~(\ref{eq:vareq}) are shown in the right panel. Different symbols correspond to different nuclear interactions  as detailed in the legend in the left panel. Solid, dashed, and dotted lines show partial contributions from the collisions between the species of same kind, different kinds, and the primed terms, respectively. With the dash-dotted lines in the panel (b) we show the effective inverse mean free paths for the electromagnetic collisions of protons (with e, $\mu$, and p) for three values of temperature, as indicated in the plot. The temperatures are (from the top dash-dotted line  to the bottom one) $T=10^7$, $10^8$, and $10^9$~K. 
}\label{fig:lk_xp015}
\end{figure*}

We solved the Bethe-Goldstone equation in the partial-wave basis up to total momentum $J=12$ on a grid of total baryon number densities spanning from $n_B=0.05$ to 0.6~fm$^{-3}$ and proton fractions from $x_p=0$ to $0.5$.

Here we analyze the same NN interaction interactions as in our previous studies \cite{Baldo2014PhRvC,Shternin2017JPhCS}. Namely, we use two realistic two-body potentials: the Argonne v18 (Av18 for short) potential \cite{Wiringa1995PhRvC} and the  charge-dependent Bonn (CD-Bonn for short) potential \cite{Machleidt2001PhRvC}. It is well-known that the non-relativistic two-body interactions fail to reproduce the saturation point of the symmetric nuclear matter. To this end it is necessary to introduce three-body nuclear forces (tbf). Here, as in the Ref.~\cite{Baldo2014PhRvC}, we include three-body forces  as an effective two-body interaction. This effective two-body interaction is obtained from the three-body one by averaging over the third particle, see Refs.~\cite{Grange1989PhRvC, Zuo2002NuPhA} for details. 

We investigate two models for the tbf. The first one is the phenomenological Urbana IX (UIX for short) model \cite{Carlson1983NuPhA}. This model contains adjustable parameters that were tuned to approach the correct saturation point with the Av18 or CDBonn two-body potentials. The second tbf model in our study is  the microscopic three-body force  (TBFmic for short) model based on the meson-nucleon theory of the nucleon interaction \cite{Li2008PhRvC,Li2012PhRvC}. The TBFmic model investigated here is based on the same meson-nucleon coupling parameters as the Av18 potential, so we use \new{it only} in combination with this two-body potential, see Ref.~\cite{Baldo2014PhRvC} for more details. In total, we show below the results for five different NN interactions: Av18, CDBonn, Av18+UIX, CDBonn+UIX, and Av18+TBFmic. 

The $G$-matrices calculated from the solution of Equation~(\ref{eq:BBG}) are taken on-shell  and on the Fermi surface (so that   the  starting energy omitted in Equation~(\ref{eq:Q_L}) \new{for brevity} is $z=e_i(p_{Fi})+e_j(p_{Fj})$) and are substituted into Equations~(\ref{eq:Q_legendre})--(\ref{eq:Q_L}).  The effective masses in the Bruecker-Hartree-Fock approach for the interactions studied here were obtained in Ref.~\cite{Baldo2014PhRvC} that provided the convenient analytical approximations for these quantities. Having $Q_{ij}$ and $m^*_{i,j}$ in hand, we can calculate the effective mean free paths and transport coefficients of nucleons in NS cores. 

\section{Results and discussion}\label{sec:discuss}
\subsection{Effective mean free paths}\label{sec:mfp}
In Figure~\ref{fig:lk_xp015} we plot the partial inverse effective mean free paths (Equation~(\ref{eq:lambda_partial})) for the thermal conductivity problem for neutrons (Figure~\ref{fig:lk_xp015}a) and protons (Figure~\ref{fig:lk_xp015}b) as a function of baryon density $n_B$ for a fixed value of the proton fraction $x_p=0.15$. As detailed in Section~\ref{sec:formalism},  the partial mean free paths mediated by NN interactions scale as $T^{-2}$, so we plot combinations $\left(\lambda^\kappa\right)^{-1} T^{-2}_8$, where $T_8\equiv T/(10^8~\mathrm{K})$ Solid, dashed, and dotted lines correspond to $\left(\lambda^\kappa_{nn}\right)^{-1}$, $\left(\lambda^\kappa_{np}\right)^{-1}$, and $\left|{\lambda'}^\kappa_{pn}\right|^{-1}$ in the left panel, respectively, and to $\left(\lambda^\kappa_{pp}\right)^{-1}$, $\left(\lambda^\kappa_{pn}\right)^{-1}$, and $\left|{\lambda'}^\kappa_{np}\right|^{-1}$, respectively, in the right panel. Notice that we plot $\left|{\lambda'}^\kappa_{pn}\right|^{-1}$ in the left, neutron,  panel, while in the right, proton, panel we show $\left|{\lambda'}^\kappa_{np}\right|^{-1}$. This is because these primed quantities couple with the neutron or proton effective mean free paths, respectively, in the system of equations (\ref{eq:vareq}). 
Different symbols correspond to different interactions considered in this paper, as indicated in the legends. 
The prominent discontinuities in $\left({\lambda'}^{\kappa}_{pn}\right)^{-1}$ and  $\left({\lambda'}^{\kappa}_{np}\right)^{-1}$ at $n_B\sim 0.45$~fm$^{-3}$ for the Av18+TBFmic interaction (dotted lines with open triangles) are manifestations of the change of sign of these quantities around this density. Notice, that these  non-diagonal elements of the scattering matrix in Equation~(\ref{eq:vareq})  can have any sign. In the conditions of Figure~\ref{fig:lk_xp015} these coefficients are negative except for Av18+TBFmic case at $n_B\gtrsim 0.45$~fm$^{-3}$.\new{We could not isolate the specific physical reason for the sign change for the particular Av18+TBFmic interaction.} In  Figure~\ref{fig:lk_xp015}b we also plot with dash-dotted lines the partial inverse mean free path for protons  $\left(\lambda^{\kappa, \mathrm{em}}_{p}\right)^{-1}$ due to the electromagnetic interaction with all charged particles, leptons (electrons and muons) and also the protons themselves.\footnote{In the latter case we neglect for simplicity the interference term between the strong and electromagnetic parts of the proton-proton interaction.} The  mean free paths mediated by electromagnetic interactions are calculated following Refs.~\cite{ShterninYakovlev2007,ShterninYakovlev2008} and obey non-Fermi liquid temperature dependence due to dynamical character of the plasma screening in the relativistic degenerate plasma \cite{Heiselberg1992NuPhA,Heiselberg:1993cr,Schmitt2018}. Therefore, $\left(\lambda^{\kappa, \mathrm{em}}_p\right)^{-1}$ is given for three temperatures, $T=10^7,\,10^8$, and $10^9$~K. In the leading order, $\lambda^{\kappa, \mathrm{em}}_p\propto T^{-1}$ instead of the standard Fermi-liquid dependence. Notice, that $\lambda^{\kappa, \mathrm{em}}_p$  depends on the proton effective mass and hence on the NN interaction. In order to not overcrowd the plot even more, in Figure~\ref{fig:lk_xp015} $\left(\lambda^{\kappa, \mathrm{em}}_p\right)^{-1}$ is shown for the Av18+UIX NN interaction only. 

\begin{figure*}[t]
\begin{minipage}{0.48\textwidth}
\includegraphics[width=\textwidth]{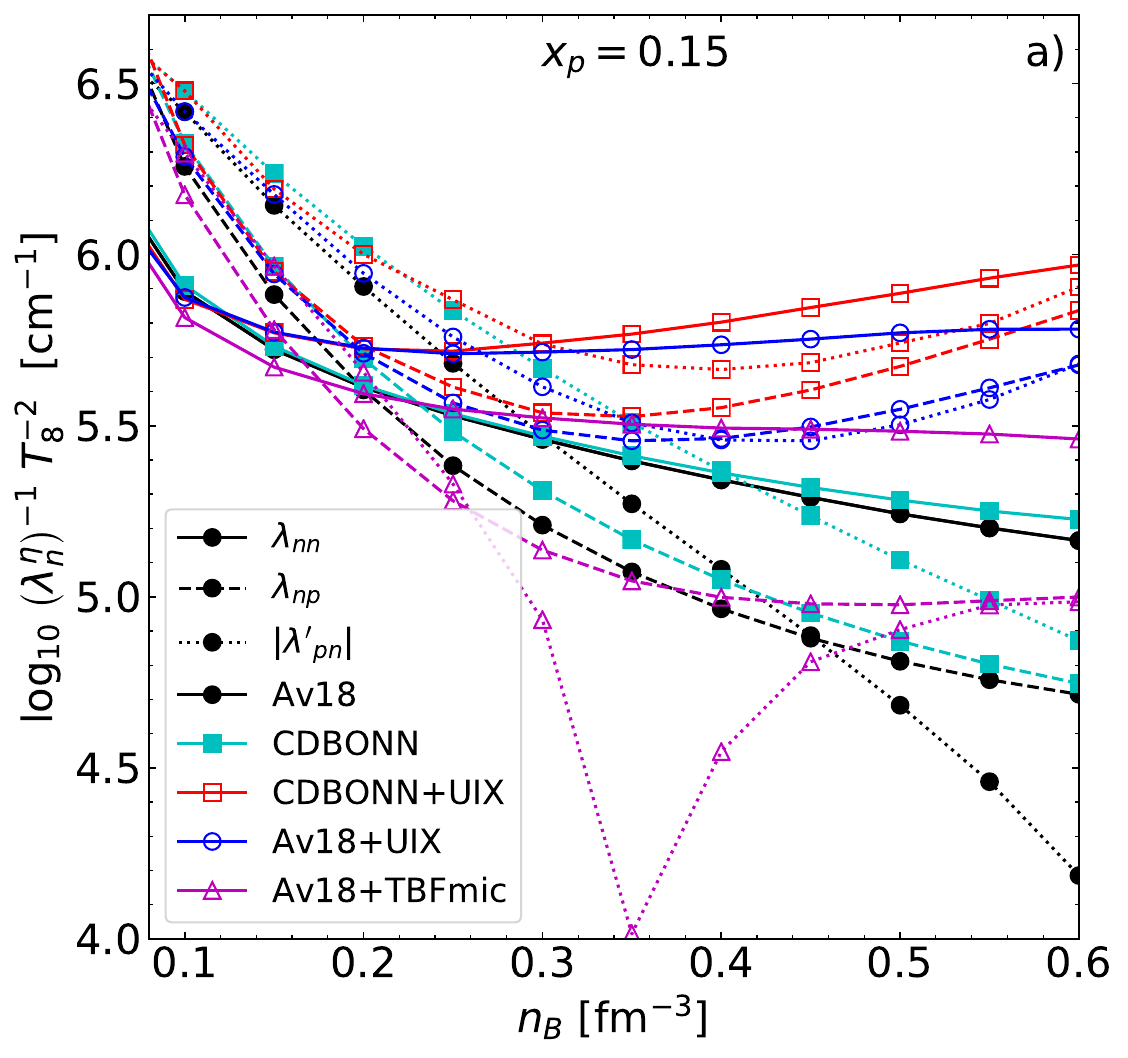}
\end{minipage}
\hspace{0.02\textwidth}
\begin{minipage}{0.48\textwidth}
\includegraphics[width=\textwidth]{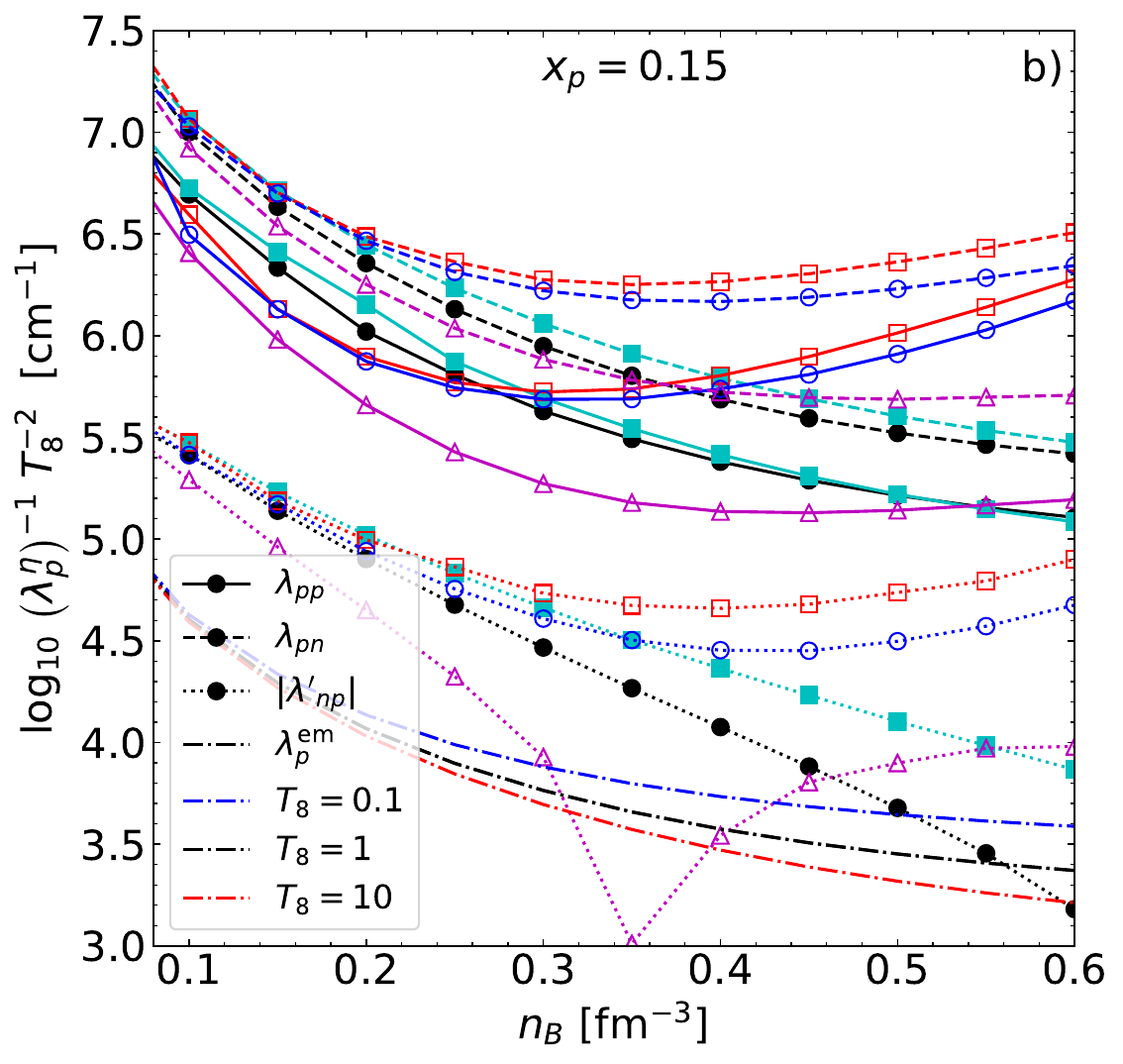}
\end{minipage}
\caption{Effective inverse mean free paths for the shear viscosity problem as functions of density for the proton fraction $x_p=0.15$. Curves and notations are the same as in Fig.~\ref{fig:lk_xp015}.}\label{fig:lh_xp015}
\end{figure*}

The total effective mean free path is limited by the most frequent collisions. In Figure~\ref{fig:lk_xp015} this corresponds to largest partial inverse mean free paths. Despite differences between the interactions used, one can conclude that at lower densities, $n_B\lesssim 0.3$~fm$^{-3}$, the neutron-proton scattering (dashed lines in Figure~\ref{fig:lk_xp015}a) dominates the neutron mean free path for the thermal conductivity problem. The reason for this originates in a larger np cross-section due to inclusion of the $I=0$ isospin channel and smaller characteristic 
c.m.
energy for the np scattering in comparison to the nn one \cite{FlowersItoh1979ApJ,Baiko2001AA,Shternin2013PhRvC}. 
 
The similar situation is observed for  protons (Figure~\ref{fig:lk_xp015}b), where the np contribution is always larger than the pp one among the strong interaction scattering channels. 
The strong interaction part of the proton-proton scattering contributes less to the total proton friction in comparison to how the neutron-neutron scattering contributes to the neutron mean free path because of the small proton fraction. In the case of protons, however, 
the electromagnetic interaction can play an important role, especially at low temperatures, see Figure~\ref{fig:lk_xp015}b.

The dependence on the choice of the NN \new{interaction} in Figure~\ref{fig:lk_xp015} is clearly seen \cite{Shternin2017JPhCS}. At low densities, $n_B\lesssim0.2$~fm$^{-3}$, inverse mean free paths calculated for different NN interactions are close, however at higher $n_B$, the results can diverge by an order of magnitude. The most prominent difference results from the selection of three-body forces. 

\begin{figure}[t]
\includegraphics[width=\columnwidth]{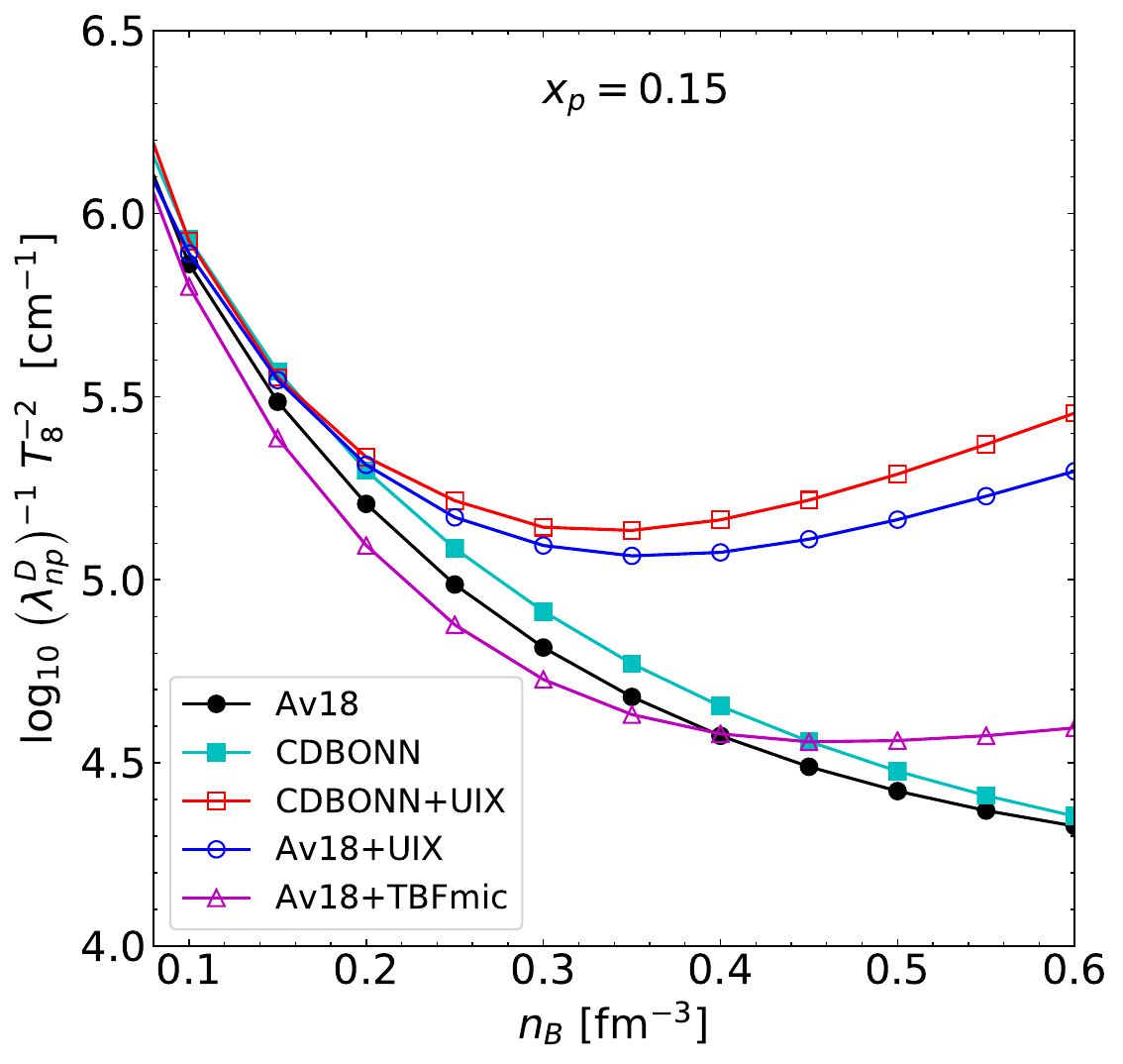}
\caption{Effective inverse mean free path for the neutron-proton scattering in the momentum relaxation problem as function of density for the proton fraction $x_p=0.15$. Different symbols correspond to different NN interactions as detailed in the legend.}\label{fig:ls_xp015}
\end{figure}
\begin{figure*}[t]
\begin{minipage}{0.48\textwidth}
\includegraphics[width=\textwidth]{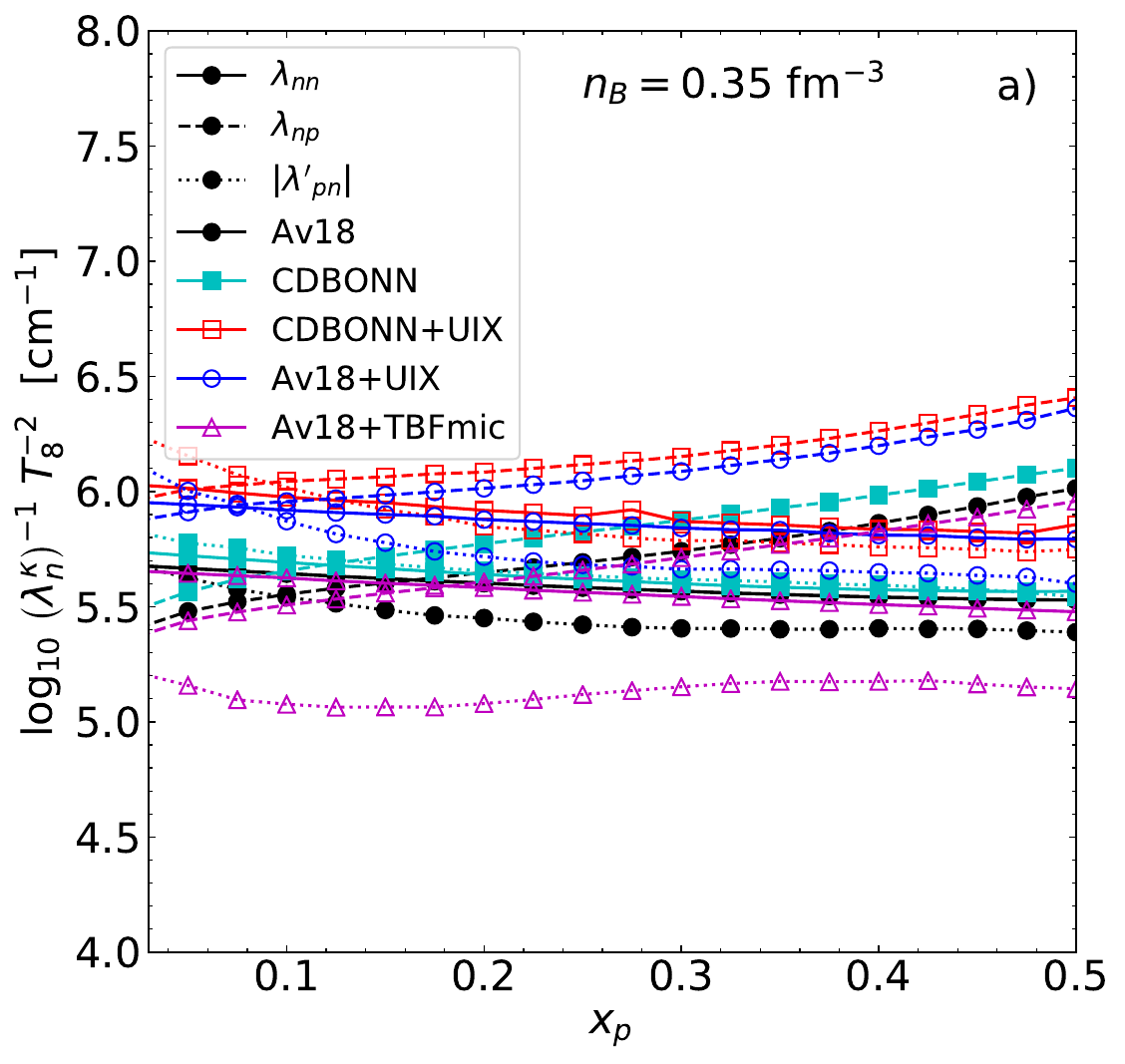}
\end{minipage}
\hspace{0.02\textwidth}
\begin{minipage}{0.48\textwidth}
\includegraphics[width=\textwidth]{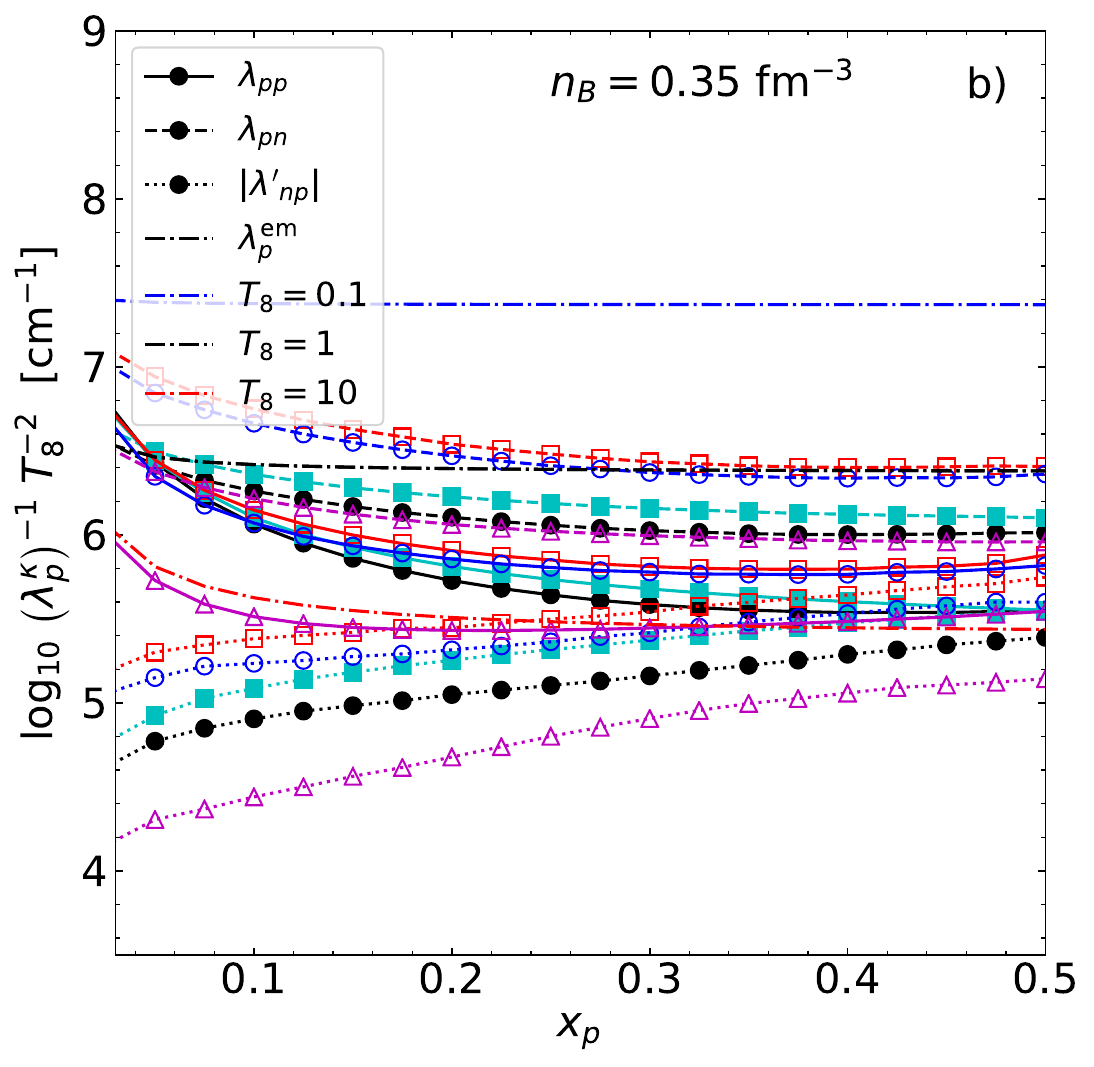}
\end{minipage}
\caption{Effective inverse mean free paths for the thermal conductivity problem as functions of the proton fraction for the baryon density $n_B=0.35$~fm$^{-3}$. Curve styles and symbols are same as in Figure~\ref{fig:lk_xp015}.}\label{fig:lk_nB035}
\end{figure*}
\begin{figure*}[ht]
\begin{minipage}{0.48\textwidth}
\includegraphics[width=\textwidth]{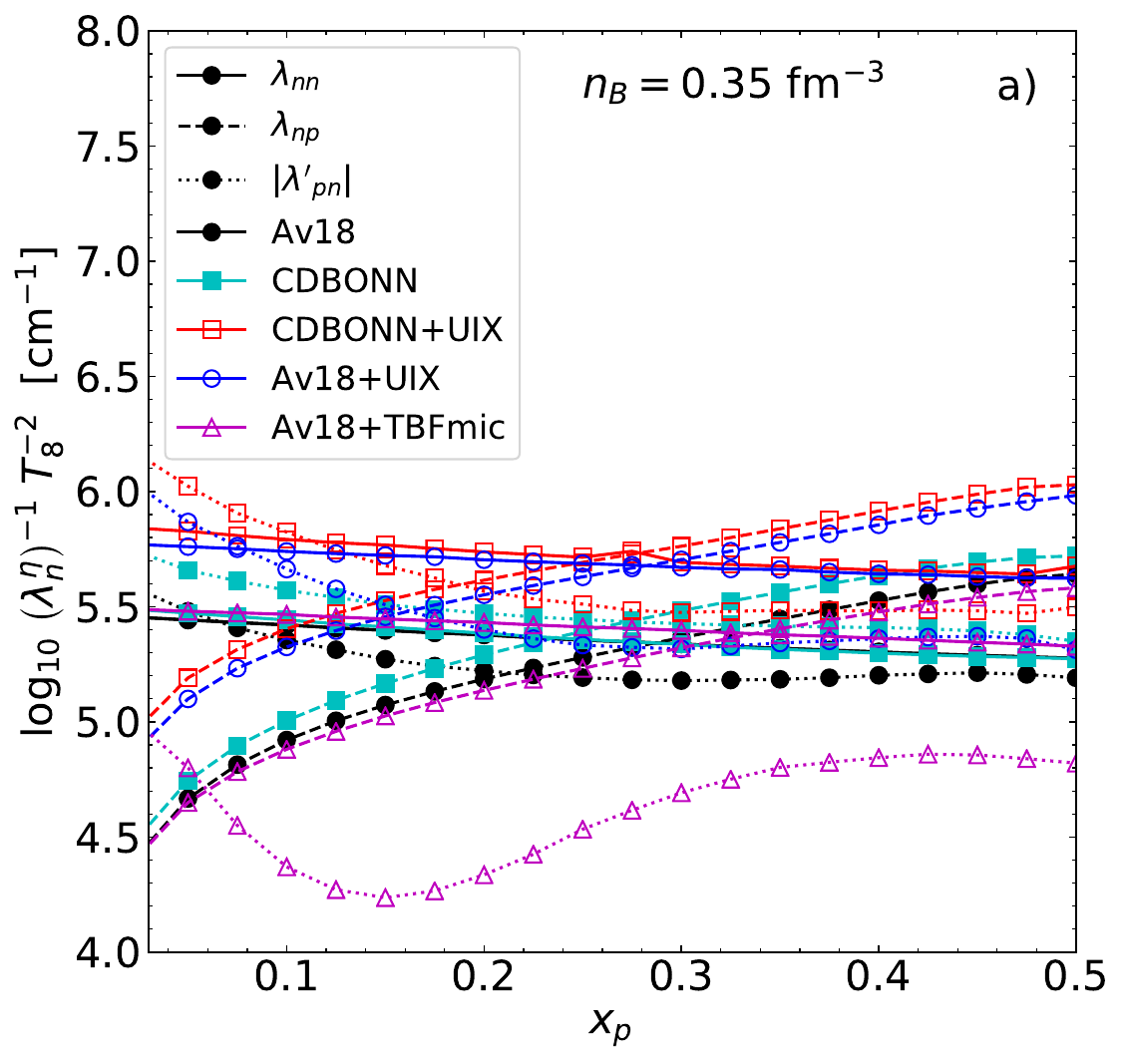}
\end{minipage}
\hspace{0.02\textwidth}
\begin{minipage}{0.48\textwidth}
\includegraphics[width=\textwidth]{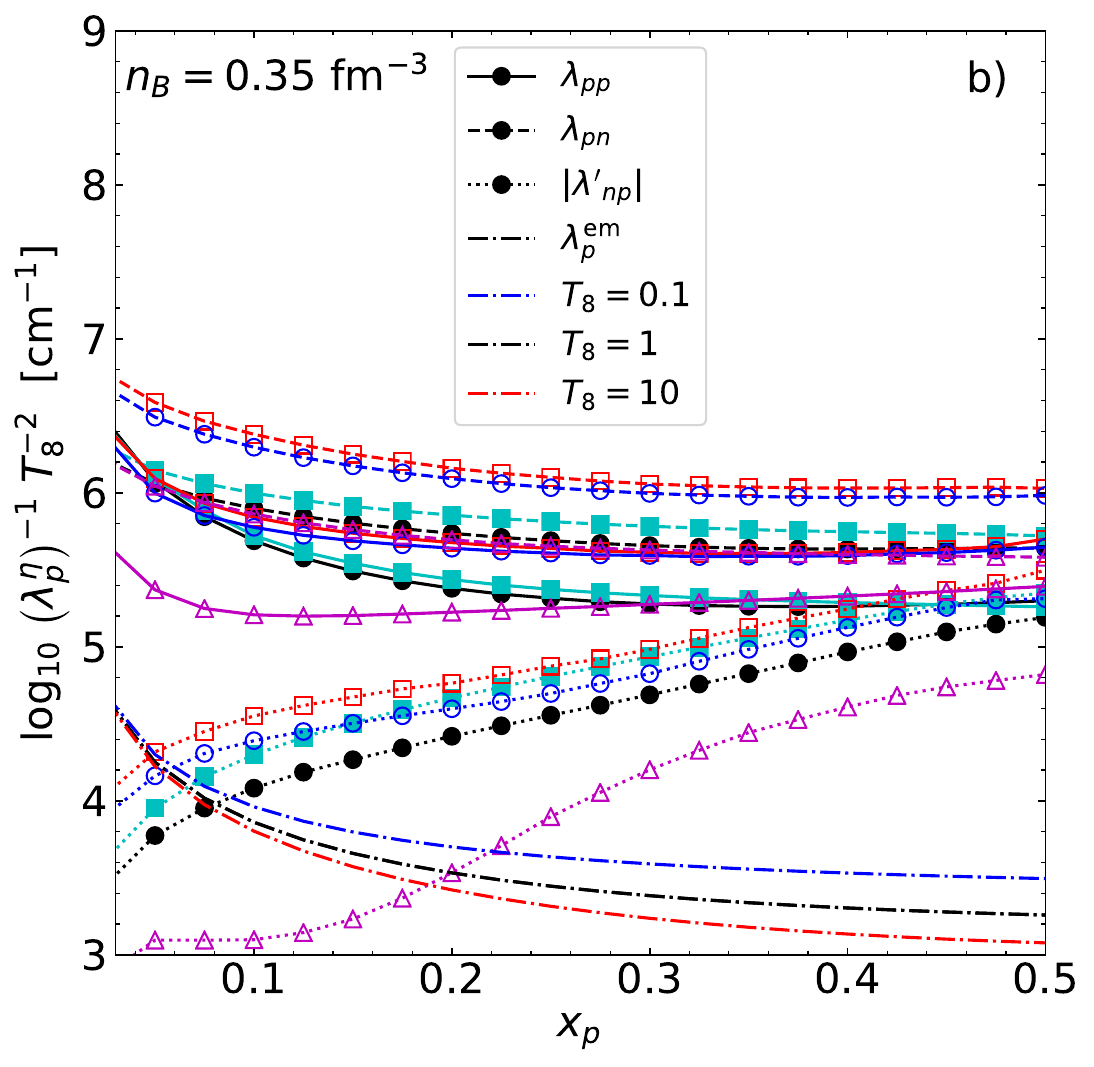}
\end{minipage}
\caption{Effective inverse mean free paths for the shear viscosity problem as functions of the proton fraction for density $n_B=0.35$~fm$^{-3}$. Curves and notations are the same as in Figure~\ref{fig:lk_xp015}.}\label{fig:lh_nB035}
\end{figure*}
\begin{figure}
\includegraphics[width=\columnwidth]{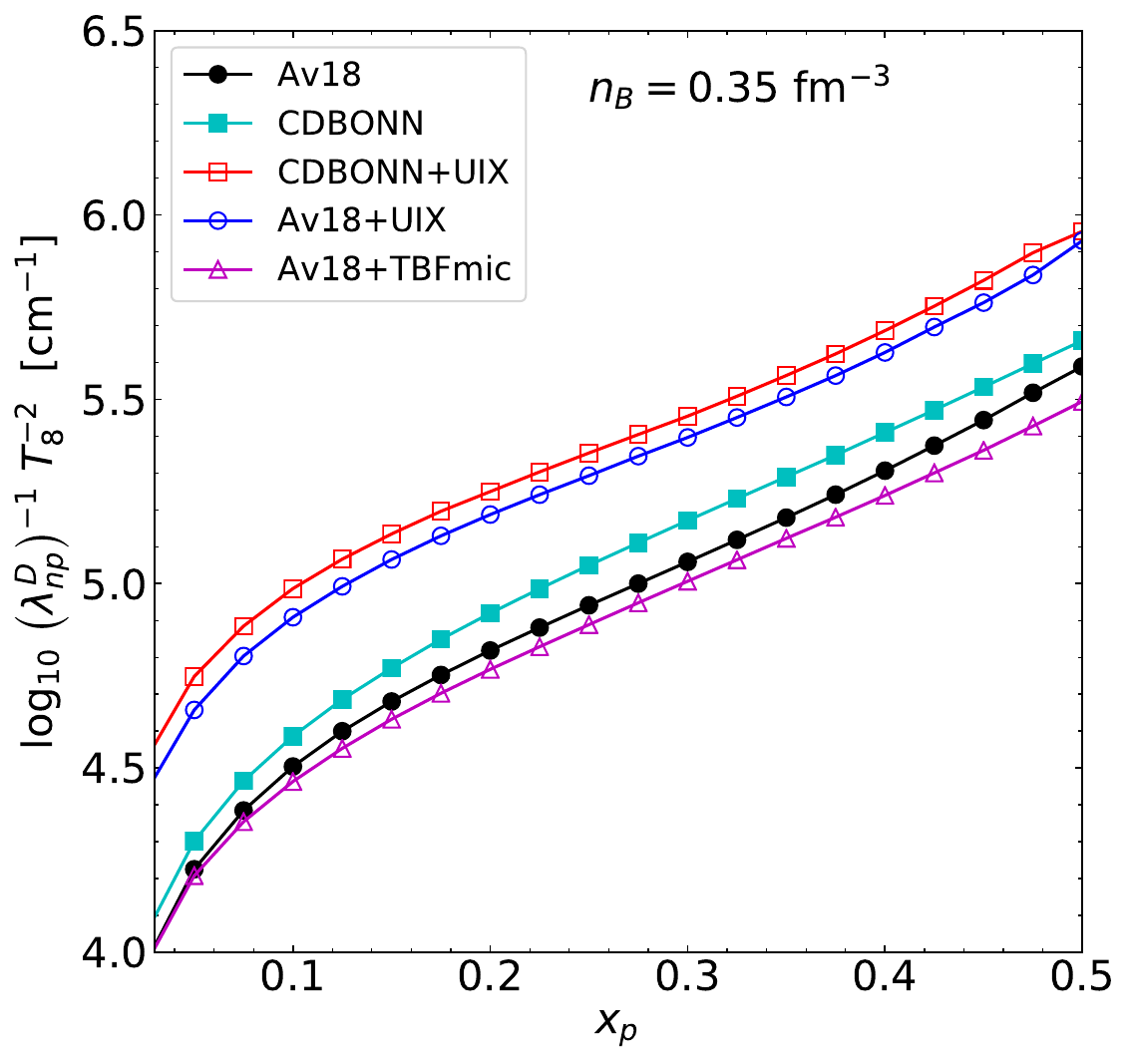}
\caption{Effective inverse mean free path for the neutron-proton scattering in the momentum relaxation problem as function of the proton fraction for the baryon density $n_B=0.35$~fm$^{-3}$. Curves and notations are the same as in Figure~\ref{fig:ls_xp015}.}\label{fig:ls_nB035}
\end{figure}
We plot similar inverse effective mean free paths for shear viscosity in Figure~\ref{fig:lh_xp015}. The design of this figure is the same as in Figure~\ref{fig:lk_xp015}. As for the case of thermal conductivity, prominent divergence of the curves representing the results for different interactions  can be observed at high $n_B$. In general, the behavior of $\left(\lambda^\eta\right)^{-1}$ is similar to $\left(\lambda^\kappa\right)^{-1}$. The qualitative difference is in the electromagnetic sector. Comparing Figures~\ref{fig:lh_xp015}b and \ref{fig:lk_xp015}b one can notice, that for the shear viscosity problem electromagnetic part of the proton interaction (dashed-dotted lines) does not contribute to the total mean free path for the protons, while for the thermal conductivity, this interaction channel can be dominant. Notice that $\lambda^{\eta, \mathrm{em}}_p$ scales as $T^{-5/3}$ (cf. $\lambda^{\kappa, \mathrm{em}}_p\propto T^{-1}$) \cite{Schmitt2018}.

In the momentum relaxation problem, only the collisions between unlike particle species contribute to the respective relaxation process. In Figure~\ref{fig:ls_xp015}, we plot the effective inverse mean free path $\left(\lambda^D_{np}\right)^{-1}$ for neutron-proton collisions which sets the momentum relaxation rate and the ambipolar diffusion timescale in the NS cores \cite{Haensel1990A&A,Goldreich1992ApJ}.

In Figures \ref{fig:lk_nB035}--\ref{fig:ls_nB035} we show the effective inverse mean free paths as  functions of $x_p$ for a baryon density $n_B=0.35$~fm$^{-3}$. Actually these values at large $x_p\gtrsim 0.3$ are not relevant in practice, since the matter inside the neutron stars is highly asymmetric. Figures \ref{fig:lk_nB035}--\ref{fig:ls_nB035} show that the dependence of the neutron-neutron mean free paths on the proton fraction is modest even at large $x_p$, while the neutron-proton scattering show considerable $x_p$ dependence. This is especially clear in Figure~\ref{fig:ls_nB035} which shows $\left(\lambda^D_{np}\right)^{-1}$.

\subsection{Approximation to the exact solution}\label{sec:exact}
Had calculated the various partial contributions to inverse mean free paths we are now in position to calculate the transport coefficients in NS cores, i.e. thermal conductivity $\kappa$, shear viscosity $\eta$, and the momentum relaxation rate $J_{np}$. 

The calculation of the latter is straightforward from the Equation~(\ref{eq:J}).
In order to calculate the former two, one needs to solve the linear system of equations (\ref{eq:vareq}). For the npe$\mu$ case considered here this is, in general, 4$\times$4 system (or 3$\times$3 system if muons are absent). The electromagnetic and nucleon sectors are coupled via protons which participate both in strong and electromagnetic interactions\footnote{Small $\ell$n collisions due to neutron magnetic moment can be neglected \cite{FlowersItoh1979ApJ,Shternin2008JETP}}. However, it turns out, that in practice the protons can be treated as the passive scatterers both for neutrons and for leptons \cite{FlowersItoh1979ApJ}. As a result, the lepton and neutron transport problems can be considered separately.
This is a consequence of a low proton fraction in the beta-stable matter of neutron star cores. 

Let us illustrate this point by considering the specific example of the Av18+UIX results at $n_B=0.35$~fm$^{-3}$, $x_p=0.15$, and $T=10^8$~K. Let us write the linear system in Equation~(\ref{eq:vareq}) as $\Lambda \bm{\lambda}=\bm{1}$, where $\bm{\lambda}$ is a vector of mean free paths, $\bm{1}$ is the right hand side vector with all components equal to 1, and $\Lambda$ is the corresponding inverse mean free paths matrix. We can write $\Lambda=\Lambda_N\oplus\Lambda_\mathrm{em}$, where $\Lambda_N$ is the 2$\times$2 (n and p) nuclear matrix which corresponds to collisions mediated by strong forces and $\Lambda_\mathrm{em}$ is the 3$\times$3 (e, $\mu$, p; or 2$\times$2 if the muons are absent) `electromagnetic' matrix which corresponds to collisions mediated by electromagnetic forces.
For the thermal conductivity problem, these matrices read
\begin{equation}
\label{eq:Lambda_kappa_N} \Lambda_{N}^\kappa=
\left(
\begin{array}{ll}
{}\quad  n &\quad p \\
\hline
 {\ }~1.67 & -0.18 \\
-0.57 & {\ }~4.2
\end{array}
\right)\times 10^6~\mathrm{cm}^{-1},
\end{equation}
\begin{equation}
\label{eq:Lambda_kappa_em}
\Lambda^\kappa_\mathrm{em}=
\left(\begin{array}{lll}
 \quad e&\quad \mu &\quad p\\
 \hline
     2.35 & 0.009 & 0.04  \\
     0.013 & 2.35 & 0.05 \\
     0.033 & 0.028 & 2.56
\end{array}
\right)
\times 10^6~\mathrm{cm}^{-1}.
\end{equation}
The non-diagonal matrix elements in $\Lambda^\kappa_{\mathrm{em}}$ are much smaller than the diagonal ones, so they can be ignored. This means that the proton mean free path does not affect the equations for the lepton  mean free paths (and vice versa). The proton-proton scattering due to nuclear forces is comparable with the proton-scattering due to electromagnetic forces (see Figures~\ref{fig:lk_xp015}b and \ref{fig:lk_nB035}b) at the selected temperature. At larger (smaller) temperatures the nuclear (electromagnetic) interactions will be limiting proton mean free path. In any case, $\lambda_p$ is smaller, and at \new{$T\lesssim 10^8$~K} significantly smaller, then one calculated from the inversion of the $\Lambda^\kappa_N$ matrix alone. Moreover $\left|{\lambda'}^{\kappa }_{np}\right|^{-1}\lambda^\kappa_p$ is always small, as Figures~\ref{fig:lk_xp015}b and \ref{fig:lk_nB035}b show. Thus the protons have a little effect on the equation for $\lambda_n$. In contrast, especially at low densities, $\left|{\lambda'}^{\kappa}_{pn}\right|^{-1}\lambda^\kappa_n$ can be large, see Figures~\ref{fig:lk_xp015}a and \ref{fig:lk_nB035}a, and $\lambda^\kappa_p$ can be strongly affected by the non-diagonal neutron term. Notice, that when important, this term $\left|{\lambda'}^{\kappa}_{pn}\right|^{-1}\lambda^\kappa_n$ is found to be negative, which lead to further suppression of $\lambda^\kappa_p$. The solution of the full system (\ref{eq:vareq}) with matrices (\ref{eq:Lambda_kappa_N})--(\ref{eq:Lambda_kappa_em}) is compared in Table~\ref{tab:kin035} with the simplified solution where protons are taken as passive scatterers. \new{That is, the off-diagonal elements in matrices (\ref{eq:Lambda_kappa_em}) and (\ref{eq:Lambda_kappa_N}) that are related to protons are set to zero.} Since proton fraction is small, and their mean free path is also small, protons always give a small contribution \new{which is not included in the simplified solution}. We also show in Table~\ref{tab:kin035} values of $\kappa_i$ for different particle species and the total thermal conductivity $\kappa$. In the case shown in Table~\ref{tab:kin035} neglecting proton contribution and selecting the simplified solution lead to error of only 10\%. 
\begin{table}[ht]
\centering
    \caption{Mean free paths for different particles for thermal conductivity and shear viscosity problems calculated from the full solution of Equation~(\ref{eq:vareq}) or the simplified solution (protons act only as scatterers), see text for details. The numbers are calculated for $n_B=0.35$~fm$^{-3}$, $x_p=0.15$, and Av18+UIX nuclear \new{interaction}. Thermal conductivity ($\kappa$) and shear viscosity ($\eta$) coefficients are also given.
    }
    \label{tab:kin035}
    \begin{ruledtabular}
    \begin{tabular}{llllll}
        &n & p &e &$\mu$ & tot\\
        \hline
    full solution &&&&\\
    \quad  $\lambda^\kappa$ ($10^{-6}$~cm) &0.62&0.20&0.42&0.42&\\
    \quad $\kappa$ ($10^{22}$~erg~cm$^{-1}$~s$^{-1}$~K$^{-1}$) &5.32 &0.53 &0.83& 0.58& 7.26\\
    \quad $\lambda^\eta$ ($10^{-6}$~cm) 
    &1.33& 0.69& 394& 256\\
    \quad $\eta$ ($10^{19}$~g~cm$^{-1}$~s$^{-1}$) &0.17&0.009&2.71&0.88&3.77\\
    simplified solution &&&&\\
    \quad  $\lambda^\kappa$ ($10^{-6}$~cm) &0.60& $-$&0.42&0.42&\\
    \quad $\kappa$ ($10^{22}$~erg~cm$^{-1}$~s$^{-1}$~g$^{-1}$) &5.14 &$-$ &0.84& 0.59 & 6.56\\
    \quad $\lambda^\eta$ ($10^{-6}$~cm) 
    &1.30& $-$& 394& 256\\
    \quad $\eta$ ($10^{19}$~g~cm$^{-1}$~s$^{-1}$) &0.17&$-$&2.71&0.88&3.76\\
     \end{tabular}
\end{ruledtabular}
\end{table}

For the shear viscosity problem, electromagnetic interaction do not affect the proton mean free path at any temperature, see Figures~\ref{fig:lh_xp015}b and \ref{fig:lh_nB035}b. Indeed, the matrices $\Lambda^\eta_N$ and $\Lambda^\eta_\mathrm{em}$ read
\begin{equation}
\label{eq:Lambda_eta_N} \Lambda_{N}^\eta=
\left(
\begin{array}{ll}
{}\quad  n &\quad p \\
\hline
 {\ }~0.77 & -0.03 \\
-0.3 & {\ }~1.9
\end{array}
\right)\times 10^6~\mathrm{cm}^{-1},
\end{equation}
\begin{equation}
\label{eq:Lambda_eta_em}
\Lambda^\eta_\mathrm{em}=
\left(\begin{array}{lll}
 \quad e&\quad \mu &\quad p\\
 \hline
     2.5 & 0.05 & 0.3  \\
     0.11 & 3.7 & 0.47 \\
     0.15 & 0.12 & 5.45
\end{array}
\right)
\times 10^3~\mathrm{cm}^{-1}.
\end{equation}
Notice a different normalization in Equation~(\ref{eq:Lambda_eta_em}). Proton mean free path due to nuclear scattering is much smaller than corresponding lepton mean free paths, therefore \new{when the lepton shear viscosity is calculated, protons can be treated as passive scatterers}. The difference of temperature scaling for $\lambda^\eta_{\mathrm{em}}$ and $\lambda^\eta_N$  is modest, therefore this conclusion holds at any temperatures.
According to Equation~(\ref{eq:Lambda_eta_N}) and Figures~\ref{fig:lh_xp015}b and \ref{fig:lh_nB035}b, $\left|{\lambda'}^{\eta}_{np}\right|^{-1}\lambda^\eta_p$ is always very small, so that the proton influence on the neutron mean free path calculation is negligible. The solution of the full system of equations in comparison to the simplified solution is also shown in Table~\ref{tab:kin035}. The relative proton contribution to the shear viscosity is much smaller than the relative proton contribution to the thermal conductivity since $\eta_p$ in Equation~\ref{eq:eta} is proportional to $x_p^{4/3}$, while $\kappa_p$ in Equation (\ref{eq:kappa}) is proportional to $x_p^{2/3}$. Therefore it is always an excellent approximation to neglect the proton contribution to the shear viscosity.

The system of equations (\ref{eq:vareq}) corresponds to the simplest variational approximation to the full system of kinetic equations. In principle, for Fermi systems it is possible to construct exact solutions of these transport equations in  form of the rapidly converging series \cite[e.g.,][]{FlowersItoh1979ApJ,Anderson1987,BaymPethick}. The closed expressions are obtained for the Fermi-liquid limit where the transition rates are energy-independent. Since the electromagnetic rates are energy-dependent due to the dynamical character of the plasma screening in the dominant interaction channel, it is not straightforward to obtain the exact solution for the full 4$\times$4 problem. However, the decoupling of the variational solutions discussed above allows on the same grounds to decouple the lepton and neutron sectors in the exact system of transport equations. In the lepton sector it turns out that the variational solution is a very good approximation to the exact solution \cite{ShterninYakovlev2007,ShterninYakovlev2008}. For the neutron transport coefficients, the calculations show that the correction to the shear viscosity coefficient do not exceed $5$\% and can be always neglected \cite{ShterninYakovlev2008,Shternin2013PhRvC,Shternin2017JPhCS}, while for the thermal conductivity this correction can be accounted for by including a factor $C_\kappa=1.2$ in Equation~(\ref{eq:kappa}) valid for all practical situations. 

To summarize, in order to calculate thermal conductivity and shear viscosity coefficients from Eqs.~(\ref{eq:kappa})--(\ref{eq:eta}) in beta-stable matter of NS cores, one needs to solve 2$\times$2 system (\ref{eq:vareq}) for e$\mu$ problem, however including the $\ell$p scattering channel. For nucleon sector it is enough to consider only neutrons scattering off neutrons and protons. The system of equations~(\ref{eq:vareq}) then reduces to only one equation for $\lambda_n$ (different for $\kappa$ and $\eta$ problems).

\subsection{Partial wave analysis}\label{sec:partial}
The results of the Section~\ref{sec:mfp} showed considerable differences in mean free paths for different nuclear interactions. It is instructive to try to understand this difference by considering different partial wave contributions.  Generally speaking, it is not easy to isolate contributions of specific partial waves, since the expression (\ref{eq:Q_L}) couples the matrix elements from the different partial waves. Still, it turns out that the dominant contribution comes from the isotropic ($L=0$) part of the scattering probability in Equation~(\ref{eq:Q_L}), with $L>0$ terms  giving less than 20\% contribution to the final transport cross-sections. This greatly simplifies an analysis. Indeed, the $L=0$ term in Equation~(\ref{eq:Q_L}) is
\begin{equation}\label{eq:QciL0}
{\cal Q}_{ij}^{(0)}=\frac{1}{16\pi^2}\sum\limits_{JS\ell\ell'} (2J+1)  \left(1+\delta_{ij}(-1)^{S+\ell}\right)^2 \left|G^{JS}_{\ell\ell'}\right|^2,
\end{equation}
being just a weighted sum of squared transition amplitudes for all available transitions between the partial wave states. 
The integration over $q$ in Equations~(\ref{eq:sigmakappa})--(\ref{eq:sigmaD}) is now trivial. Save for the normalization factor, $Q^{(0)}_{ij}$  in Equation~(\ref{eq:Q_legendre}) is a total scattering cross-section on the Fermi surface (see, e.g., the discussion in Ref.~\cite{Shternin2013PhRvC}). Thus, in this approximation, transport coefficients are determined by the total cross-section averaged over total momentum $P$ with certain weighting factor (see Appendix B in Ref.~\cite{Shternin2013PhRvC} for more details). There are several differences in how the $G$-matrix elements enter the expression for scattering versus how they contribute to the binding energy, Equation~(\ref{eq:energy_part}). First of all, in  Equation~(\ref{eq:energy_part}), only the diagonal $G$-matrix elements are involved, while the non-diagonal $\ell\neq \ell'$ elements from the coupled partial waves also equally contribute to \new{Equation~(\ref{eq:QciL0})}. Importantly, in Equation~(\ref{eq:QciL0}) all partial wave amplitudes add in squares, so both the repulsive and attractive components increase the total scattering cross-section, while they can compensate each other when combined in the total energy. Finally, the transport scattering occurs at the Fermi surface, while the whole Fermi sea of nucleons contributes to the total binding energy. In particular, this leads to appearance of the effective masses (that represent densities of states on the Fermi surface \cite{BaymPethick}) in the expressions for the transport cross-sections in Equations (\ref{eq:sigmakappa})--(\ref{eq:sigmaD}).

Figures~\ref{fig:lk_xp015}--\ref{fig:ls_nB035} show that when the two-body potential is changed from Av18 to CDBonn one, the difference in mean free path calculations is relatively small. This is reasonable, since both these potentials are from the family of so-called `realistic' potentials designed to reproduce the wealth of the experimental data. 
On the other hand, different three-body forces lead to considerably different results.  Therefore in the rest of this subsection we explore the effects of the three-body forces using only the  Av18 potential on the two-body level for brevity. 
In additions we discuss only nn and np partial mean free paths as practically relevant ones following the arguments in Section~\ref{sec:exact}.

\begin{figure}
    \centering
    \includegraphics[width=\columnwidth]{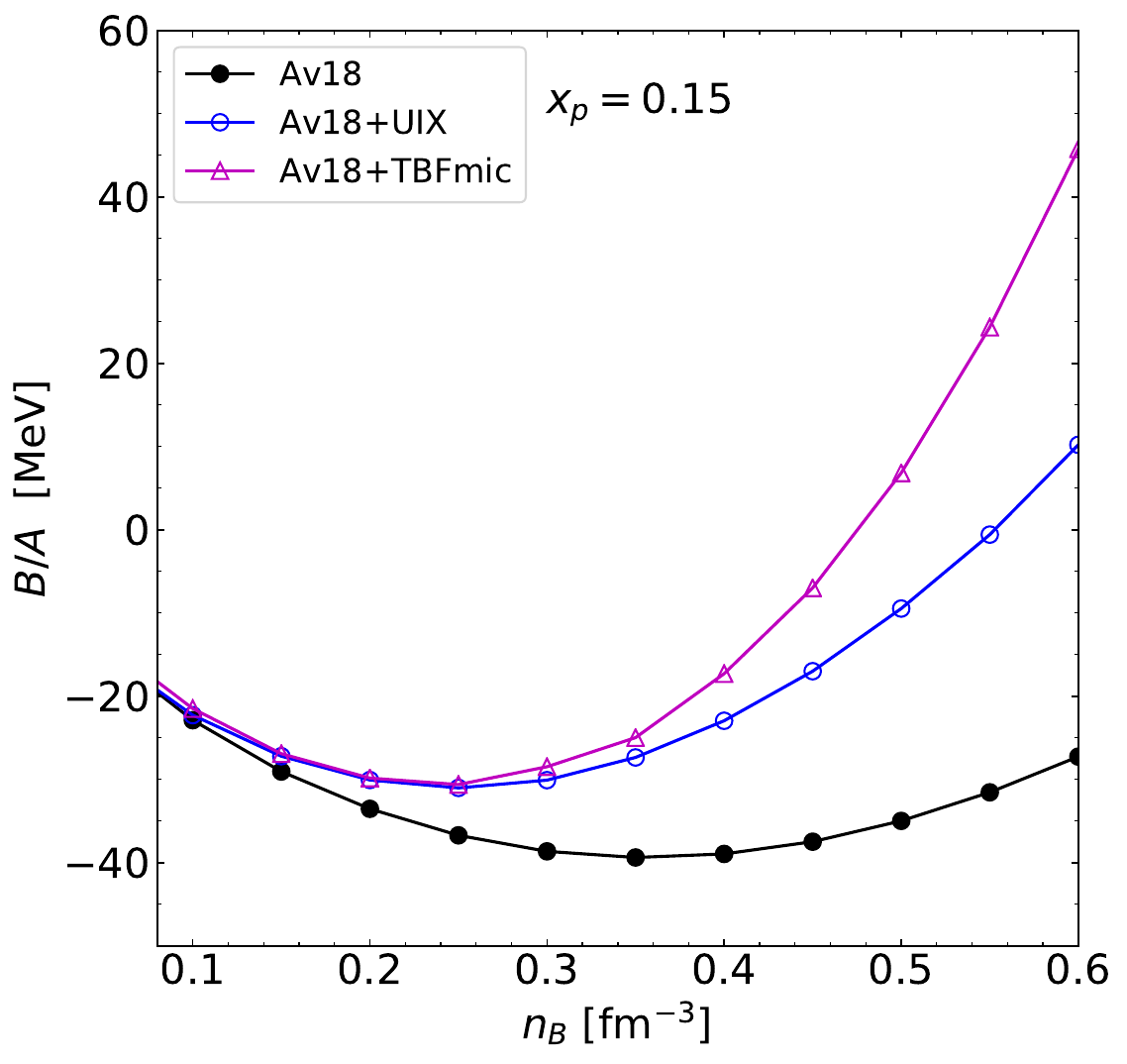}
    \caption{Binding energy per nucleon as a function of the baryon density $n_B$ for the fixed proton fraction $x_p=0.15$. The results are shown for Av18 potential at two-body level and for two three-body interaction models as indicated in the legend.}
    \label{fig:Etot}
\end{figure}
First, in Figure~\ref{fig:Etot} we show the total binding energy per nucleon, $B/A$, as a function of density for the same proton fraction $x_p=0.15$ as in Figures~\ref{fig:lk_xp015}--\ref{fig:ls_xp015} and three NN interactions, Av18, Av18+UIX, and Av18+TBFmic. The microscopic three-body force (TBFmic) is more repulsive than the UIX three-body force and its effect on the total binding energy is more prominent. However, the situation is different for scattering, as  Figures~\ref{fig:lk_xp015}--\ref{fig:ls_nB035} show. Generally, UIX tbf (open circles in Figures~\ref{fig:lk_xp015}--\ref{fig:ls_nB035}) has larger effect on scattering than the TBFmic one (open triangles in Figures~\ref{fig:lk_xp015}--\ref{fig:ls_nB035}).

\begin{figure}
    \centering
    \includegraphics[width=\columnwidth]{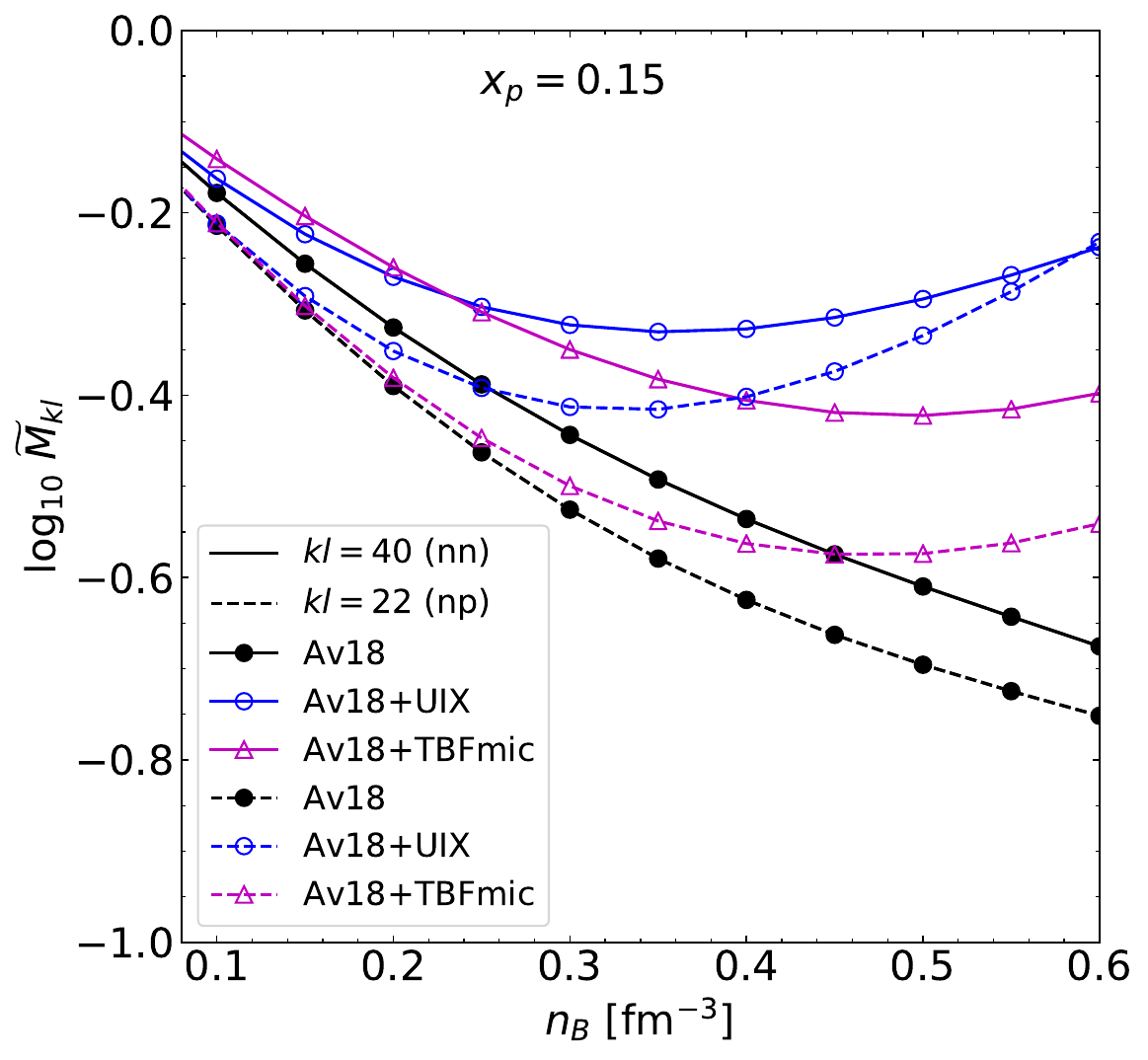}
    \caption{The factors $\widetilde{M}_{40}$ (solid lines) and $\widetilde{M}_{22}$ (dashed lines) defined in Equation~(\ref{eq:Mkl}) as functions of $n_B$ for the proton fraction $x_p=0.15$.  Different symbols correspond to different NN interactions as shown in the legend. 
    }
    \label{fig:meff_av18uix}
\end{figure}

Equations (\ref{eq:sigmakappa})--(\ref{eq:sigmaD}) show that the many-body effects enter the expressions for scattering through the squared scattering matrix elements ${\cal Q}_{ij}$ and through the effective masses $m_{i,j}^*$. The neutron-neutron and neutron-proton transport cross-sections contain different effective mass prefactors which can be commonly written as
\begin{equation}\label{eq:Mkl}
    \widetilde{M}_{kl}=m^{*k}_{n}m^{*l}_n.
\end{equation}
These are the same factors as introduced for the neutrino emission processes in Ref.~\cite{Baldo2014PhRvC}. Specifically, the neutron-neutron transport cross-sections contain $\widetilde{M}_{40}$ prefactor, similarly to the neutron-neutron bremsstrahlung \cite{Baldo2014PhRvC}, while the neutron-proton transport cross-sections, like the neutron-proton bremsstrahlung rate, contain the $\widetilde{M}_{22}$ prefactor \cite{Baldo2014PhRvC}. It is not clear how to isolate the contribution of different partial waves to the nucleon effective masses. Therefore we only consider the overall $\widetilde{M}_{kl}$ effect. For the Av18+UIX interaction the effective masses are larger than for Av18 or Av18+TBFmic interactions \cite{Baldo2014PhRvC} contributing partly to the differences in the results shown in  Figures~\ref{fig:lk_xp015}--\ref{fig:ls_nB035}. In Figure~\ref{fig:meff_av18uix} we plot the factor $\widetilde{M}_{40}$ appropriate for the neutron-neutron scattering with solid lines. These results show that the effective masses are responsible for about 0.2dex difference between the Av18+TBFmic and Av18+UIX results and 0.5dex between Av18 and Av18+UIX results. 
Similar results are found for the 
factor $\widetilde{M}_{22}$ appropriate for the neutron-proton scattering. These factors are shown for three different NN interactions in Figure~\ref{fig:meff_av18uix} with dashed lines. At large densities, effective masses are responsible for a factor of 3 difference between the Av18+UIX results and Av18 results for np scattering, and for a factor of 2 difference between Av18+UIX results and Av18+TBFmic results.

Effective masses are only partially responsible for the differences between the results obtained for different NN interactions. Another part of the difference comes from the scattering matrix elements. Below we discuss this second contribution for neutron-neutron and neutron-proton scattering, taking the effective mass prefactors $\widetilde{M}_{kl}$ out.

\begin{figure}
    \centering
    \includegraphics[width=\columnwidth]{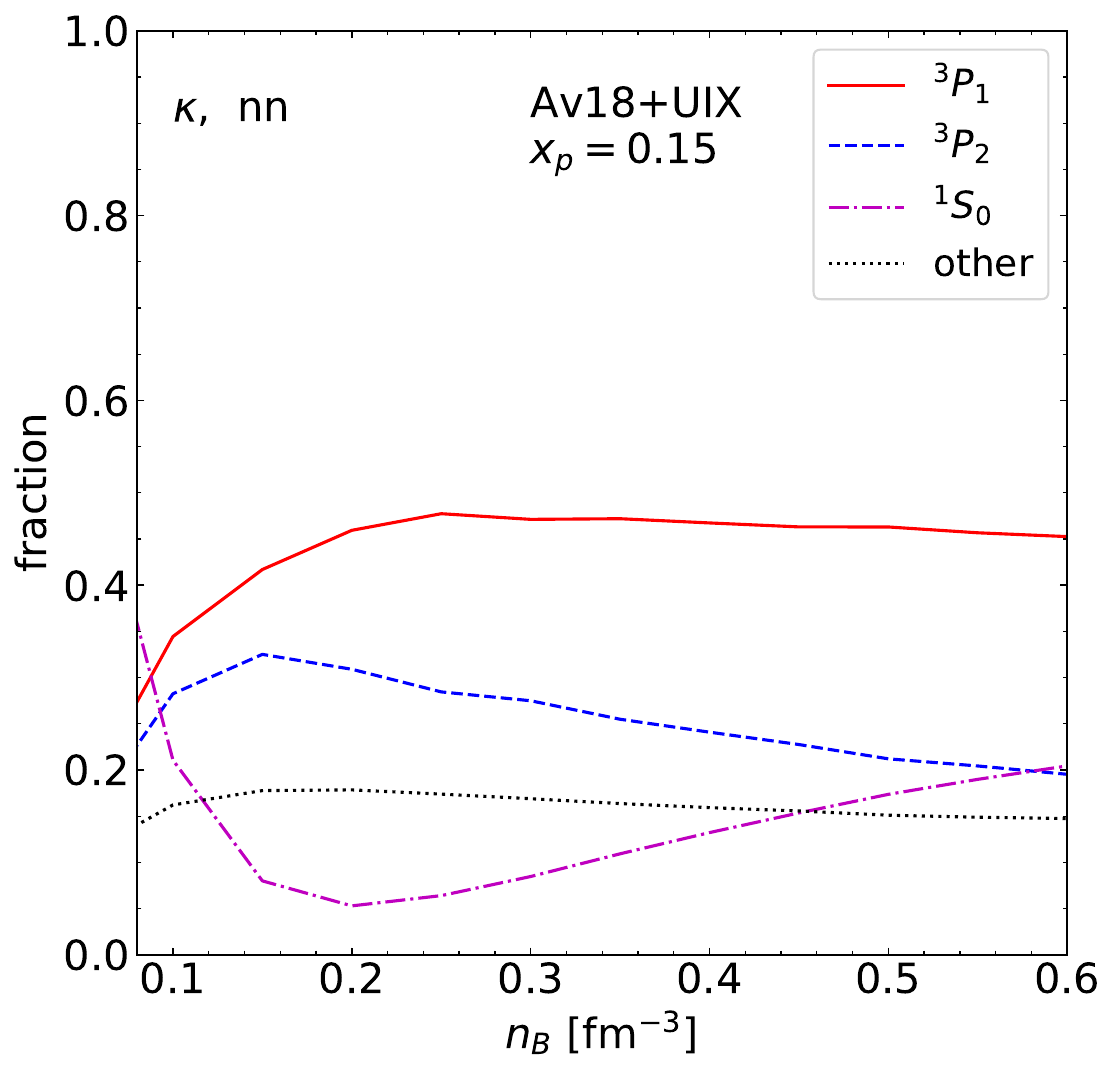}
    \caption{Relative contributions of different partial wave channels to the $L=0$ part of the nn scattering inverse mean free path $\left(\widetilde{\lambda}^\kappa_{nn}\right)^{-1}_0$ for the thermal conductivity problem defined in Equation~(\ref{eq:lambda_kappann0}) as  functions of $n_B$ for $x_p=0.15$. Results for the Av18+UIX interaction are shown.  Solid, dashed, and dash-dotted lines correspond to the ${}^3P_1$, ${}^3P_2$, and ${}^1S_0$ channels, respectively. Dotted line shows the cumulative contribution from all other channels.}
    \label{fig:lpart_av18uix}
\end{figure}

Consider first the neutron-neutron scattering. We will show the specific results for the 
thermal conductivity problem only. The results for the shear viscosity problem are similar. 
As follows from Equations~(\ref{eq:lambda_partial}), (\ref{eq:sigmakappa}), (\ref{eq:sigmakappaprime}), and (\ref{eq:Q_legendre}), the contribution from the isotropic part of the transport cross-section, $Q^{(0)}_{nn}$, to the inverse nn mean free path is
\begin{equation}\label{eq:lambda_kappann0}
\left(\widetilde{\lambda}^\kappa_{nn}\right)_{0}^{-1}\equiv \left(\lambda^\kappa_{nn} \widetilde{M}_{40}\right)^{-1}_{0} = \frac{T^2}{10\pi^2 p_{Fn}^4}\left\langle {\cal Q}_{nn}^{(0)}(4p_{Fn}^2-P^2)\right\rangle,
\end{equation}
where $\widetilde{M}_{40}$ factor gets rid of the effective masses and the subscript index 0 indicates that only $L=0$ contribution is included.
Our analysis shows that the dominant contributions 
to this quantity comes from three partial wave channels, namely ${}^3P_1$, ${}^3P_2$, and ${}^1S_0$.   Figure~\ref{fig:lpart_av18uix} shows the relative contributions 
of these partial wave channels 
to $\left(\widetilde{\lambda}^\kappa_{nn}\right)^{-1}_{0}$ as function of $n_B$ for $x_p=0.15$. For concreteness, the Figure~\ref{fig:lpart_av18uix} is plotted for the Av18+UIX interaction. For other interactions one observes qualitatively similar situation, although the proportion between different channels can vary. Still, these three channels are dominant for any NN interaction analyzed in this paper. The dominant channel is the  ${}^3P_1$ one, which relative contribution is shown with the solid line in Figure~\ref{fig:lpart_av18uix}. At very small densities, the ${}^1S_0$ channel dominates (dash-dotted line in Figure~\ref{fig:lpart_av18uix}), but its contribution at intermediate densities decreases, and the ${}^3P_2$  channel (dashed line) is more important. The ${}^1S_0$ channel, which matrix element increases considerably at small $p$, is additionally suppressed by the angular weighting factor $4p_{Fn}^2-P^2=4p^2$ in Equation~(\ref{eq:lambda_kappann0}). Al large densities, the ${}^1S_0$ channel for the UIX tbf  becomes important again.
At the two-body level, both ${}^1S_0$ and ${}^3P_2$ contributions are actually less important at high density. The rest of the contribution to $\left(\widetilde{\lambda}^\kappa_{nn}\right)^{-1}_{0}$ comes from other partial waves as shown with the dotted line in Figure~\ref{fig:lpart_av18uix}. Their individual contributions are small and can not be resolved. In total, they are of the same order of magnitude as the contribution form the higher, $L>0$, mutlipole moments in Equations~(\ref{eq:Q_legendre})--(\ref{eq:Q_L}).

\begin{figure}
    \centering
    \includegraphics[width=\columnwidth]{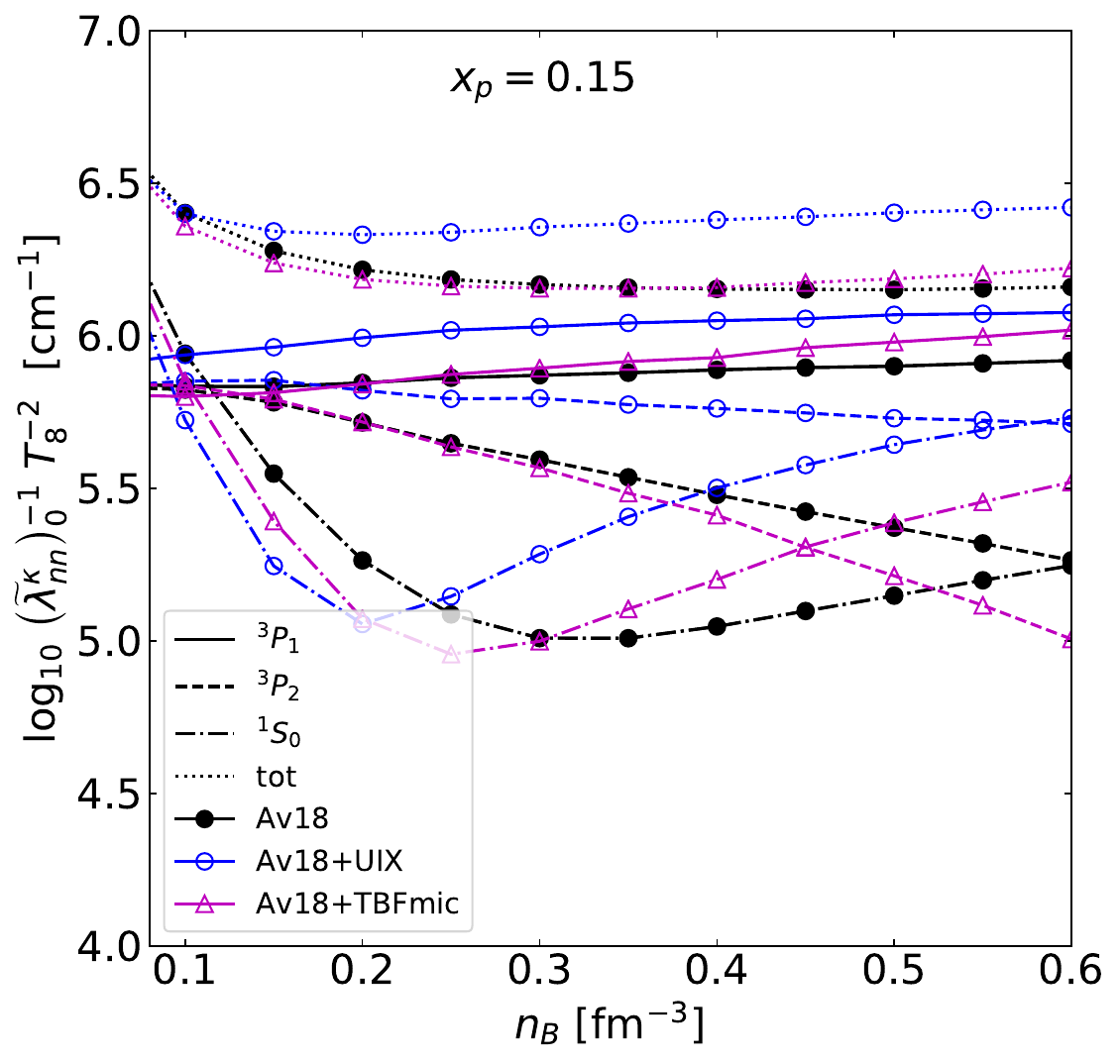}
    \caption{Partial contributions of different partial wave channels to the $L=0$ part of the nn scattering inverse mean free path $\left(\widetilde{\lambda}^\kappa_{nn}\right)^{-1}_0$ for the thermal conductivity problem as  functions of $n_B$ for $x_p=0.15$. Effective mass prefactor is not included, see Equation~(\ref{eq:lambda_kappann0}). Solid, dashed, and dash-dotted lines correspond to the ${}^3P_1$, ${}^3P_2$, and ${}^1S_0$ channels, respectively. Dotted line is the total contribution to $L=0$ part, including the partial waves not shown in the figure.
    Different symbols correspond to different interactions.}
    \label{fig:lpart}
\end{figure}

In Figure~\ref{fig:lpart} we compare the contributions to $\left(\widetilde{\lambda}^\kappa_{nn}\right)^{-1}_{0}$ from three dominant partial waves in absolute values for the Av18 (filled circles), Av18+UIX (open circles), and Av18+TBFmic (open triangles) NN interactions. In addition, with dotted lines we plot the total inverse mean free path $\left(\widetilde{\lambda}^\kappa_{nn}\right)^{-1}_{0}$.
As before, the results are plotted for $x_p=0.15$. The UIX tbf contains more repulsion in the ${}^3P_1$ channel (solid lines) than the microscopic three-body force, which translates to somewhat larger values for the inverse mean free path in this channel for the Av18+UIX case. 
Additionally, TBFmic is much less attractive in the ${}^3P_2$ channel than UIX tbf, so the absolute value of the contribution is larger for UIX force and it gives larger contribution to scattering. In general, the UIX tbf has larger (in magnitude) contributions both in the repulsive and attractive low-angular momenta interaction channels. When sum up to the total energy, the positive and negative contributions partially cancel and the TBFmic force have larger binding energy (Figure~\ref{fig:Etot}). However in scattering the squared absolute values of the partial wave matrix elements are important and the Av18+UIX interaction results in largest scattering cross-sections among the NN interactions considered here.

\begin{figure}
    \centering
    \includegraphics[width=\columnwidth]{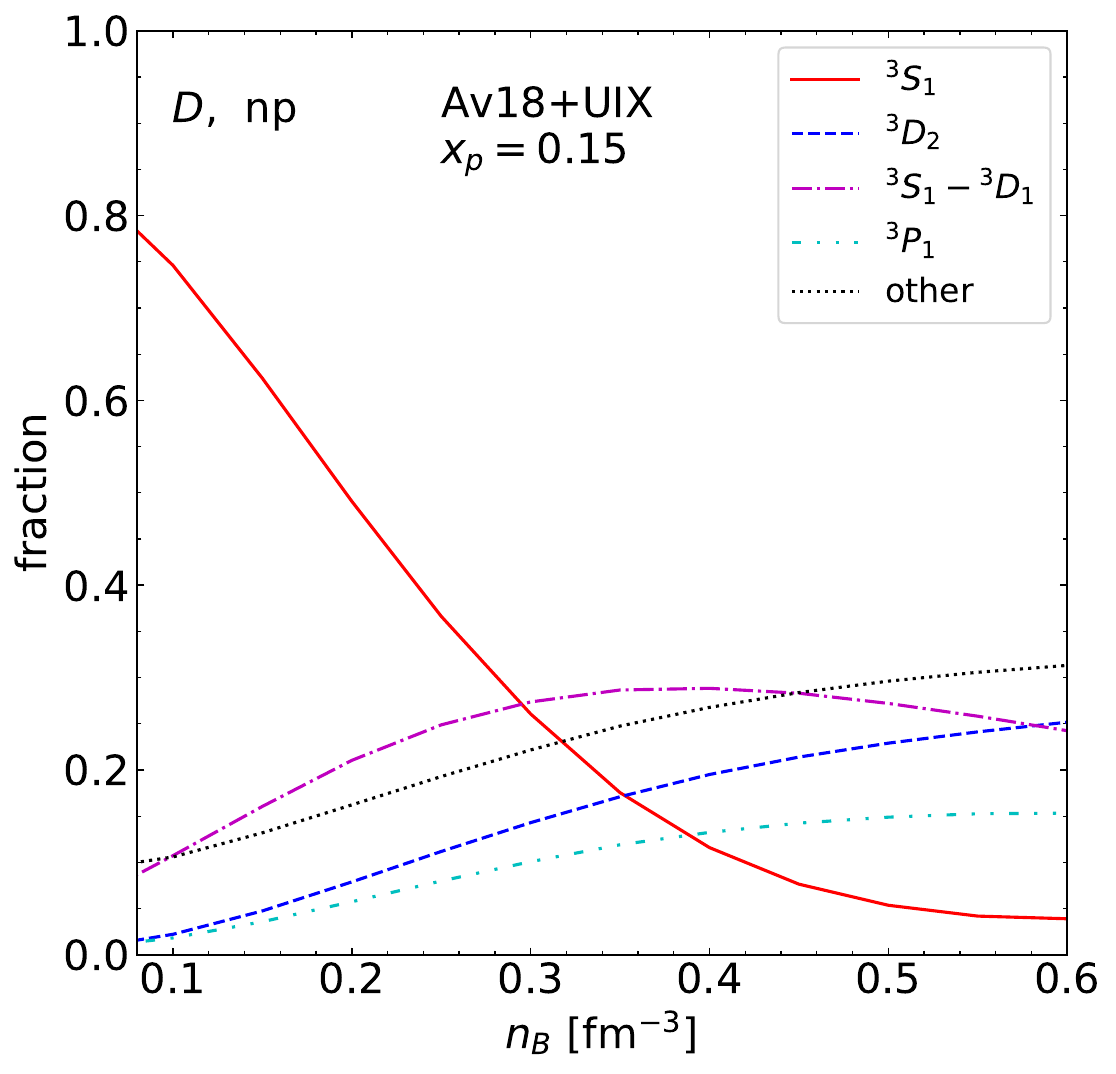}
    \caption{Relative contributions of different partial wave channels to the $L=0$ part of the np scattering inverse mean free path $\left(\widetilde{\lambda}^D_{nn}\right)^{-1}_{0}$ for the momentum relaxation problem defined in Equation~(\ref{eq:lambda_sigmaD0}) as functions of $n_B$ for $x_p=0.15$. Results for the Av18+UIX interaction are shown.  Solid, dashed, dash-dotted, and dash-double dotted lines correspond to the ${}^3S_1$, ${}^3D_2$, ${}^3S_1-{}^3D_1$, and ${}^3P_1$ channels, respectively. Dotted line shows the cumulative contribution from all other channels.}
    \label{fig:lsigmapart_av18uix}
\end{figure}
\begin{figure}[th]
    \centering
    \includegraphics[width=\columnwidth]{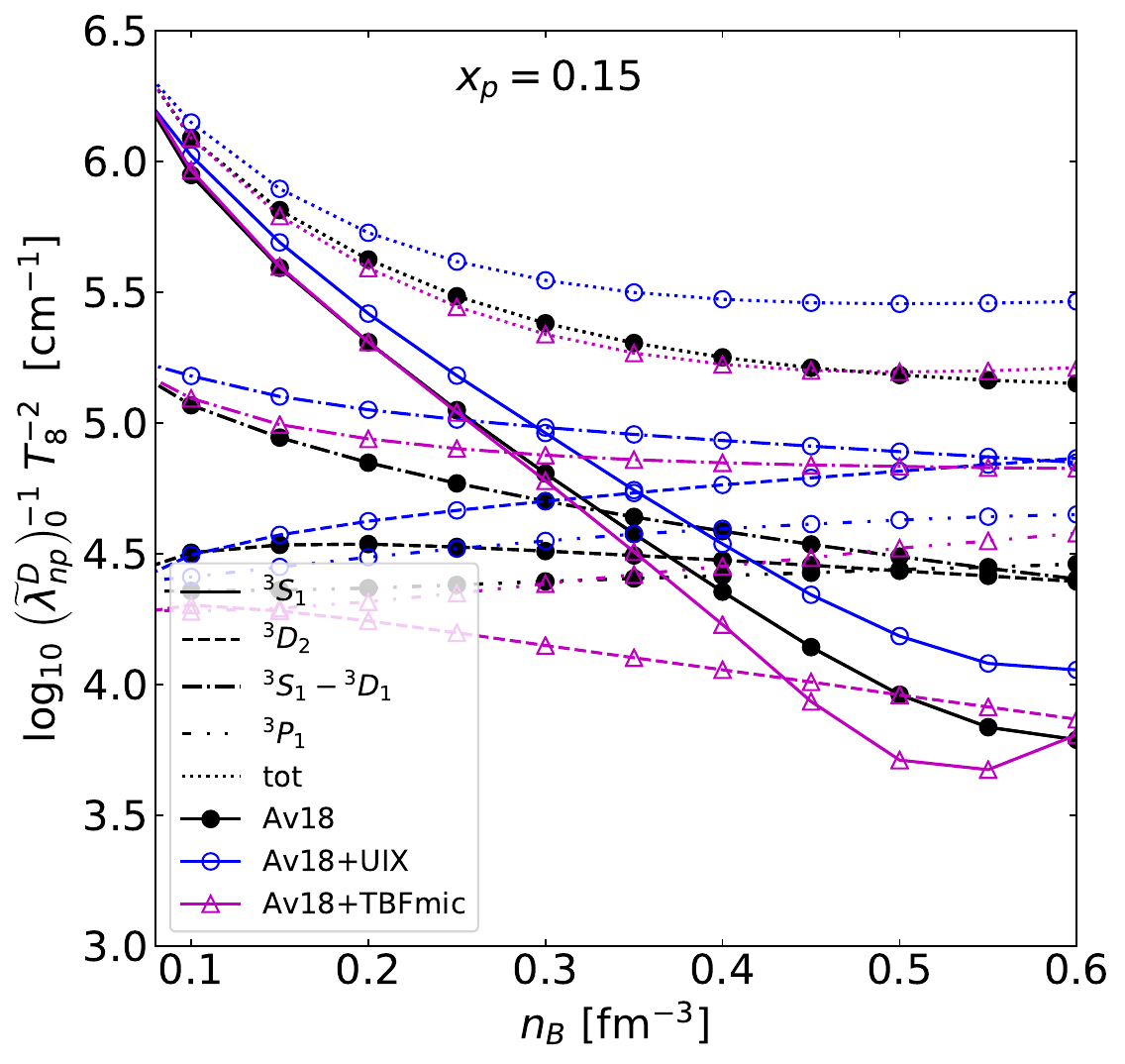}
    \caption{Partial contributions of different partial wave channels to the $L=0$ part of the np scattering inverse mean free path $\left(\widetilde{\lambda}^D_{np}\right)^{-1}_0$ for the thermal conductivity problem as functions of $n_B$ for $x_p=0.15$. Solid, dashed, and dash-dotted lines correspond to the ${}^3P_1$, ${}^3P_2$, and ${}^1S_0$ channels, respectively. Different symbols correspond to different interactions.}
    \label{fig:lsigmapart}
\end{figure}
The analysis of the neutron-proton scattering is less transparent. 
Here we chose the momentum relaxation problem as an example on np scattering, Equation~(\ref{eq:sigmaD}). The analysis of the neutron-proton scattering for the thermal conductivity or shear viscosity shows the same results.\footnote{The difference is in the  angular factors in Equations~(\ref{eq:sigmakappa}), (\ref{eq:sigmaeta}), and (\ref{eq:sigmaD}) which does not change qualitatively the relative contributions from different partial waves}
In analogy with Equation~(\ref{eq:lambda_kappann0}), using Equations~(\ref{eq:sigmaD}) and (\ref{eq:Q_legendre}) we define
\begin{equation}\label{eq:lambda_sigmaD0}
    \left(\widetilde{\lambda}^D_{np}\right)^{-1}_{0}\equiv\left(\lambda^D_{np} \widetilde{M}_{22}\right)^{-1}_{0} = \frac{T^2}{6\pi^2 p_{Fn}^4}\left\langle {\cal Q}_{np}^{(0)} q^2 \right\rangle.
\end{equation}
The  dominant contribution to $\left(\widetilde{\lambda}^D_{np}\right)^{-1}_0$ at low densities is the s-wave ${}^3S_1$ channel. However at larger densities many channels give comparable contributions and it is hard to isolate a single or a few dominant terms. Nevertheless, in Figure~\ref{fig:lsigmapart_av18uix} we show, in analogy to Figure~\ref{fig:lpart_av18uix}, the relative contributions of a few partial waves to $\left(\widetilde{\lambda}^D_{np}\right)^{-1}_0$. The next important contribution after ${}^3S_1$ channel comes from the non-diagonal term $\ell'=\ell+2$ in Equation~(\ref{eq:QciL0}) representing the ${}^3S_1-{}^3D_1$ coupling. The fractional contribution of this channel is shown with dash-dotted line in Figure~\ref{fig:lsigmapart_av18uix}. Remember that the non-diagonal terms do not contribute to the total energy.
Also important contribution at high density comes from the ${}^3D_2$ channel (dashed line in Figure~\ref{fig:lsigmapart_av18uix}) and, less significantly, ${}^3P_1$ channel (double-dot-dashed line in Figure~\ref{fig:lsigmapart_av18uix}) that was dominant in the nn scattering. We do not separate the rest of the partial waves, and show their total contribution by dotted line in Figure~\ref{fig:lsigmapart_av18uix}. At high densities it is as much as about 30\%, however individual contributions are less than 5\% thus it is not relevant to discuss them separately. Among them, the next channels in order of importance are  ${}^1P_1$ and ${}^3D_2$, although this is density- and interaction-dependent.

In Figure \ref{fig:lsigmapart} we plot four main partial wave contributions to $\left(\widetilde{\lambda}^D_{np}\right)^{-1}_0$. One can see that the main contribution, which results in stronger scattering by the Av18+UIX interaction in comparison to Av18+TBFmic interaction at large $n_B$, comes from the ${}^3D_2$ partial wave (dashed lines). The scattering of the microscopic three-body force in this channel (dashed line with open triangles) is much smaller than both the Av18+UIX (dashed line with open circles) and bare two-body Av18 result (dashed line with filled circles). 
 For instance, 
 the ${}^3S_1-{}^3D_1$ contributions for Av18+UIX and Av18+TBFmic are similar. 
We can conclude that the main difference between the UIX and TBFmic results for np scattering prominent at high densities in Figures~\ref{fig:lk_xp015}--\ref{fig:ls_nB035} is partially due to difference in scattering in ${}^3D_2$ channels and partially due to larger effective masses for the UIX tbf in comparison to the  microscopic tbf \cite{Baldo2014PhRvC}.

\subsection{Practical expressions}\label{sec:pract}
\begin{figure*}[th]
\begin{minipage}{0.32\textwidth}
\includegraphics[width=\textwidth]{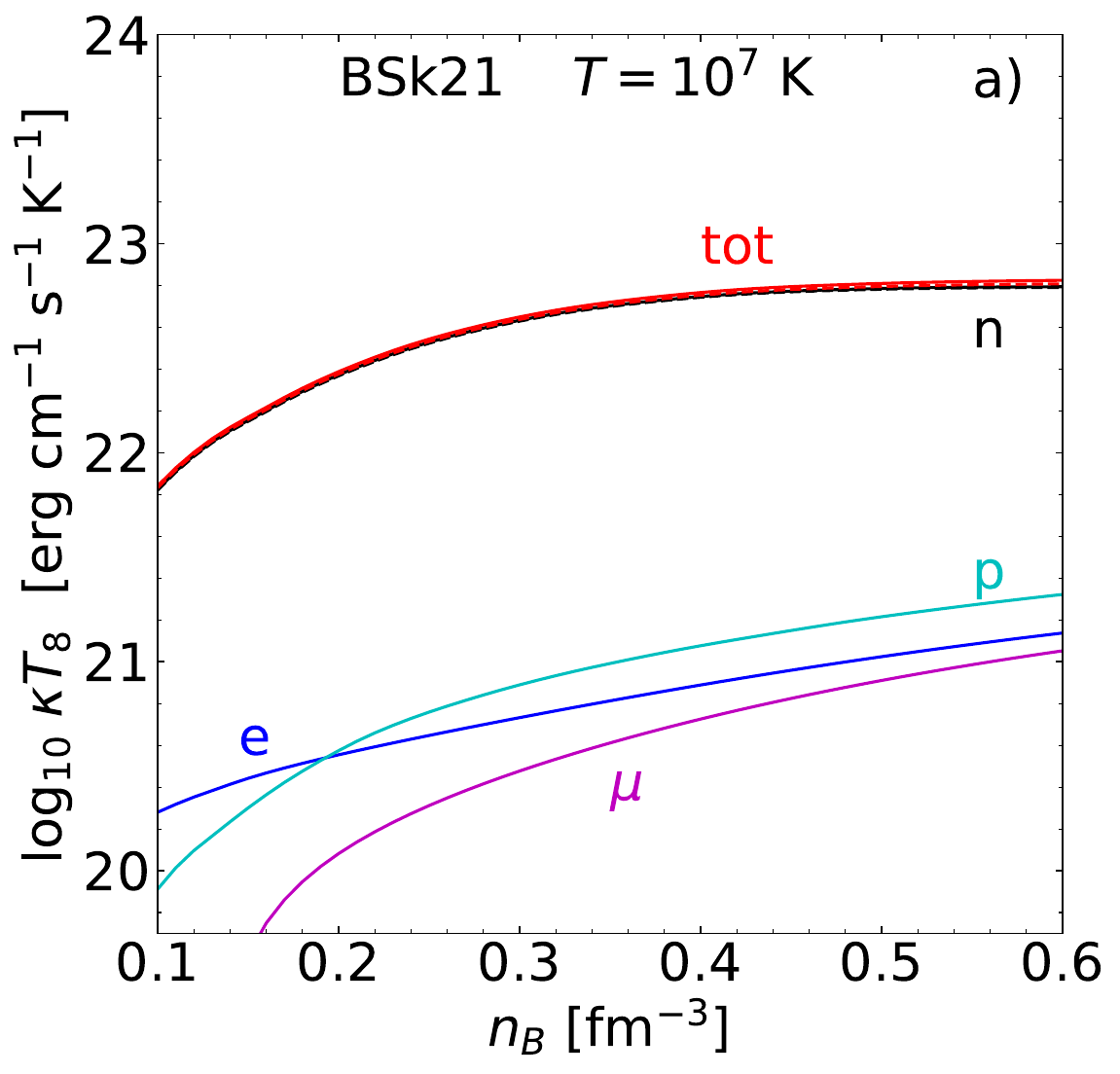}
\end{minipage}
\begin{minipage}{0.32\textwidth}
\includegraphics[width=\textwidth]{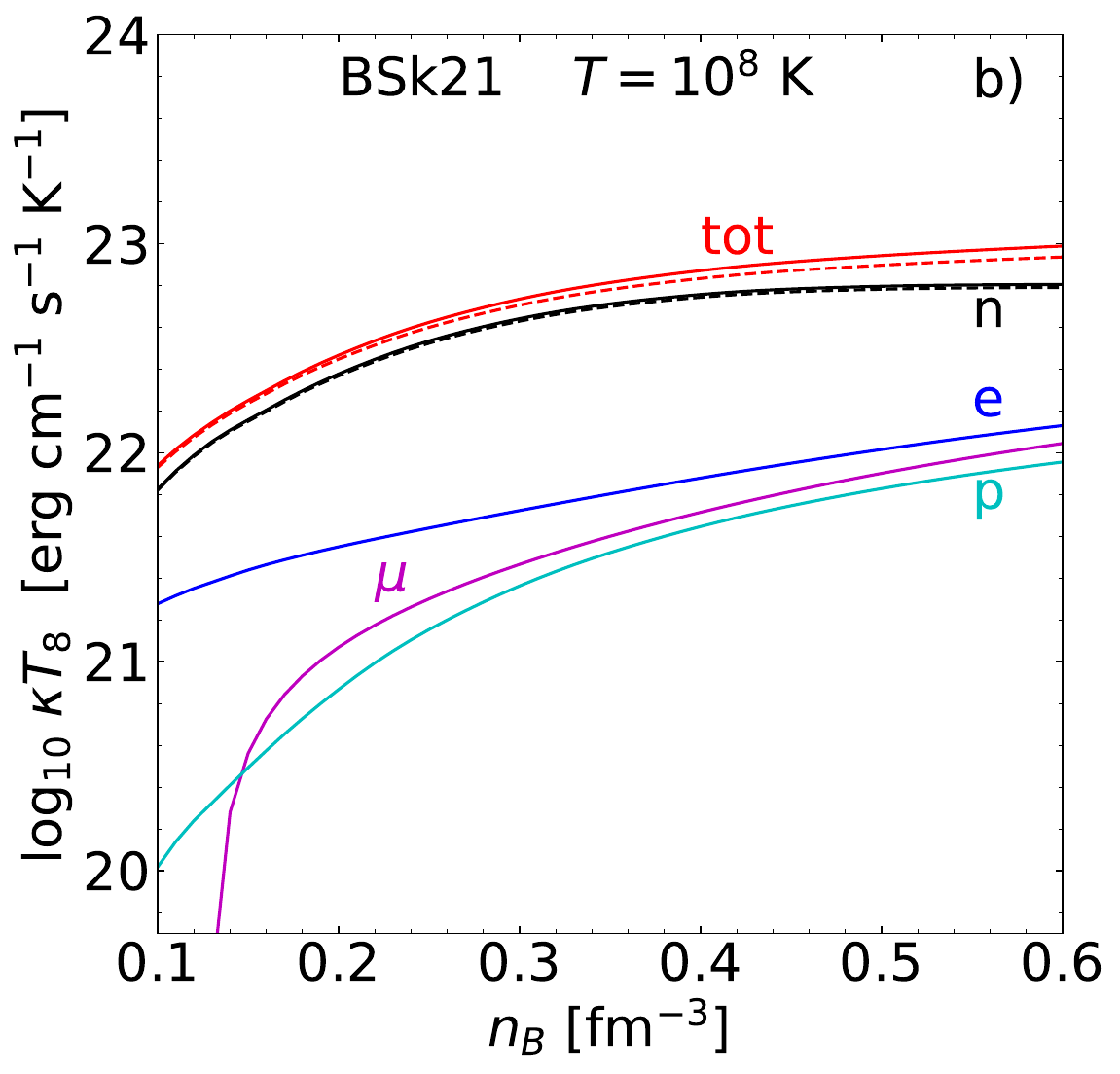}
\end{minipage}
\begin{minipage}{0.32\textwidth}
\includegraphics[width=\textwidth]{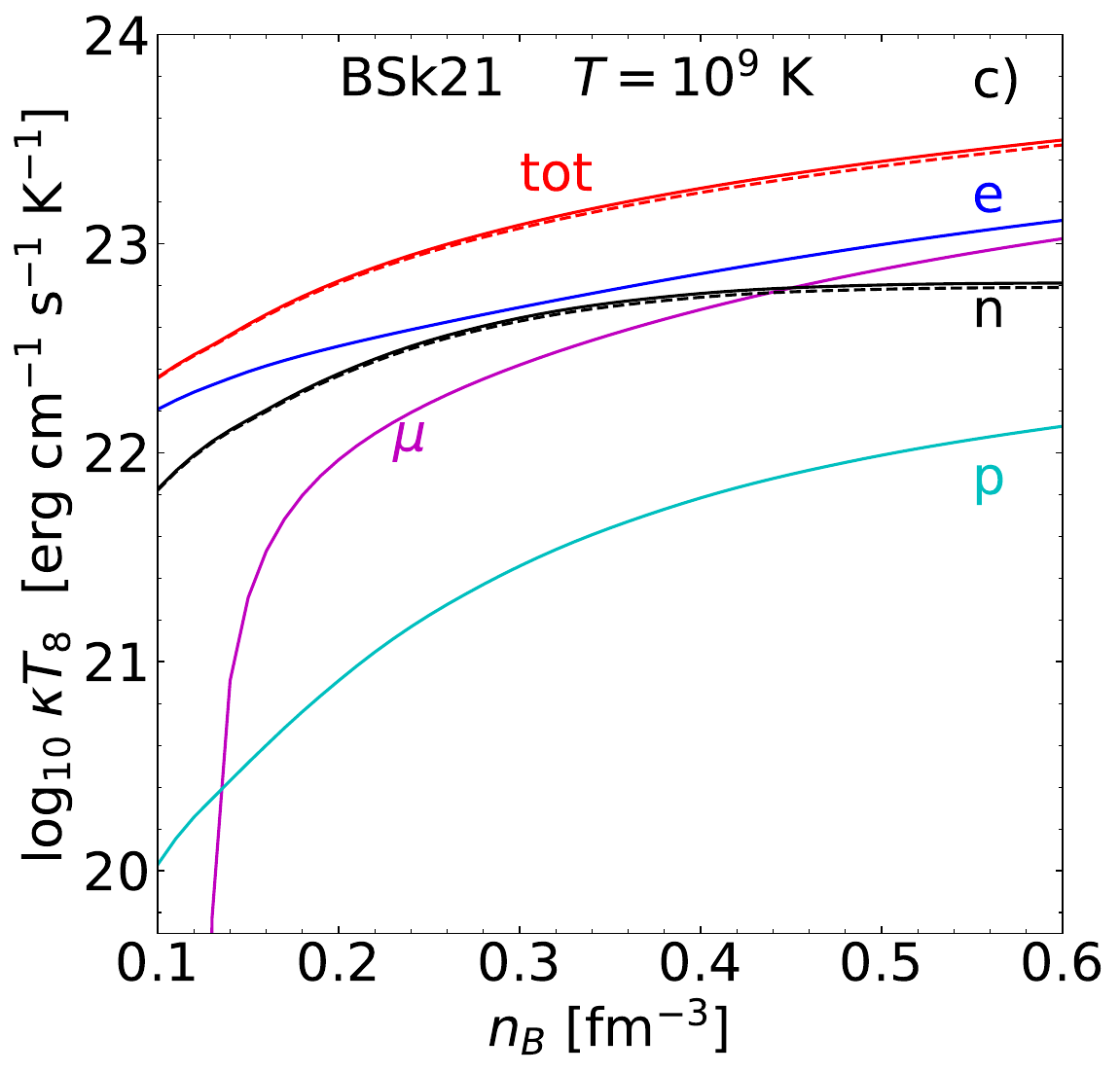}
\end{minipage}
\caption{Partial contributions to the thermal conductivity (variational solution) for the beta-stable NS core with the BSk21 equation of state and three values of temperature, $T=10^7$~K (a), $10^8$~K (b), and $10^9$~K (c). Particle species are labeled near the curves, the uppermost curves `tot' show the total thermal conductivity. Nucleon mean free paths are calculated according to the Av18+UIX model. Dashed lines corresponding to the 'tot' and 'n' curves \new{show the results of} the approximate treatment of $\kappa_n$, see text for details.  }\label{fig:kappa_BSkpart}
\end{figure*}

According to the results of the previous subsection, the mean free paths in the nucleon sector depend considerably on the selected nuclear interaction. Ideally, to perform the consistent study, the transport coefficients should be calculated based on  the same microscopic model as the EOS. In practice this is rarely possible. 
On the other hand, the variations in the results of the microscopic calculations show that it is not possible to obtain the universal expression for the transport coefficients equally applicable for any nucleon EOS of the NS core.

Here we suggest a  tradeoff approach between the consistency and universality. That is, we will take mean free paths obtained in the previous sections as a functions of $n_B$ and $x_p$ 
and use them for any EOS in question. As an example, we employ the popular BSk21 EOS based on the Brussels-Skyrme nucleon interaction functional \cite{Potekhin2013A&A}. 
Its significant advantage in practice is its fully analytical prameterization.
We consider the beta-stable matter in BSk21 NS core. The proton fraction as a function of $n_B$ is of course different than one obtains for the beta-stable matter with each of the five EOSs based on the microscopic models considered here \cite{Baldo2014PhRvC}.

We illustrate this approach in the next few Figures. In Figure~\ref{fig:kappa_BSkpart} we show partial contribution of different particle species in npe$\mu$ NS cores to the total thermal conductivity. The temperature-independent (in Fermi-liquid) combination $\kappa T_8$ is shown.
Three panels correspond to three temperatures $T=10^7$~K (left), $T=10^7$~K (middle), and $T=10^9$~K (right). The nucleon mean free paths here are calculated in the Av18+UIX model; taking the different interactions changes the picture quantitatively, but not qualitatively. 
Solid lines show the partial contributions from neutrons (n), protons (p), electrons (e), and muons ($\mu$), as labelled  near the corresponding curves. These values are calculated from the  solution of the full 4$\times$4 system of variational equations (\ref{eq:vareq}). The curves marked `tot' in each panel show the total contribution. The dashed lines barely seen in the plots correspond to the simplified approximation discussed in the Section~\ref{sec:exact}. Here e$\mu$ and n sectors are decoupled and protons are treated as passive scatterers. We see that this approximation is very good as expected. The electromagnetic scattering does not obey the Fermi-liquid behavior due to the long range of the interaction, therefore corresponding $\kappa_\ell T_8$ combinations are not temperature-independent. In the leading order, $\kappa_\ell$ is independent of $T$ \cite[e.g.,][]{Heiselberg:1993cr,ShterninYakovlev2007,Schmitt2018}. Except for the highest temperatures ($T=10^9$~K, Figure~~\ref{fig:kappa_BSkpart}c), the neutron contribution is always dominant over the lepton one and the proton contribution can be always neglected. 

\begin{figure*}[ht]
\begin{minipage}{0.32\textwidth}
\includegraphics[width=\textwidth]{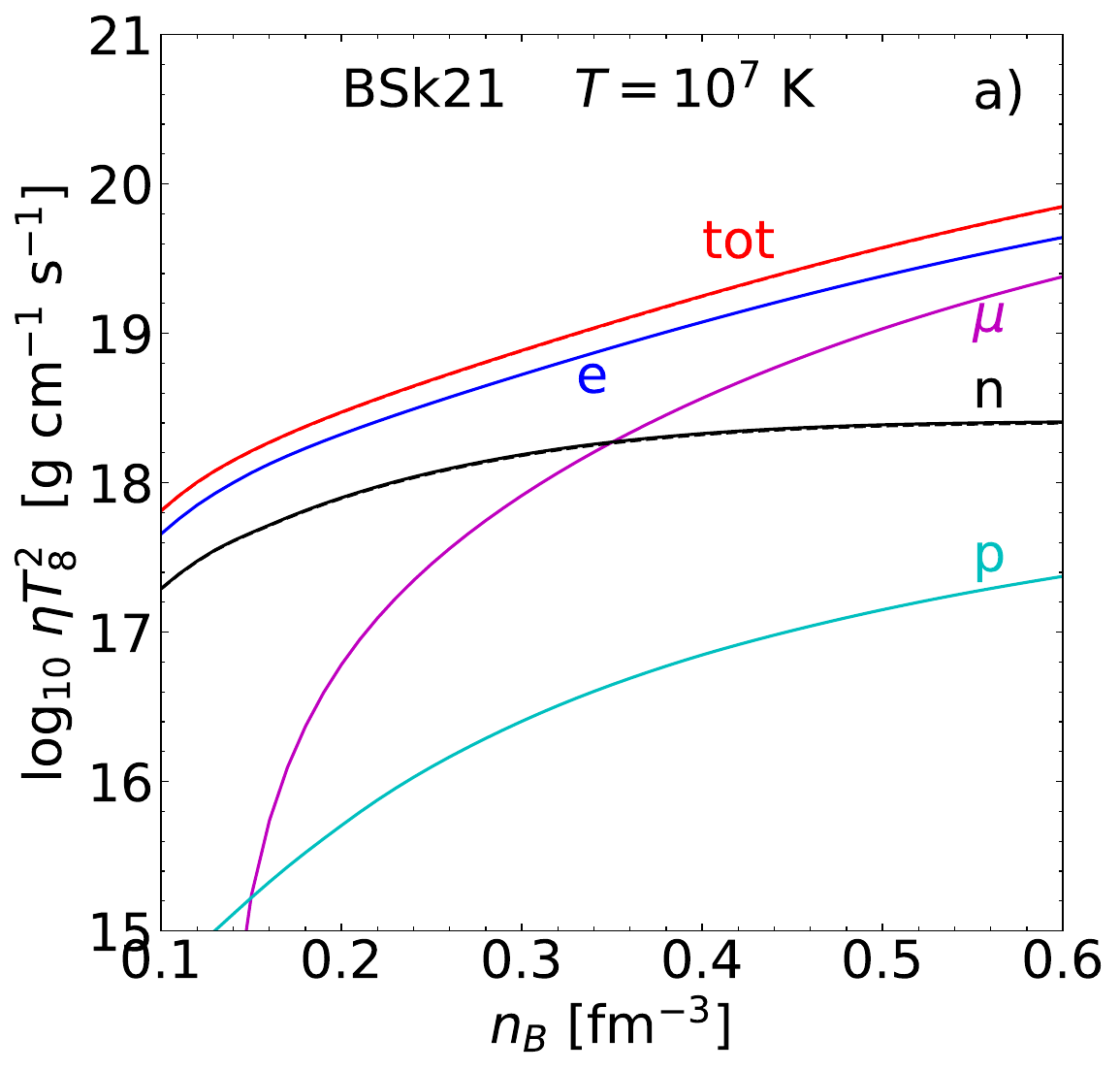}
\end{minipage}
\begin{minipage}{0.32\textwidth}
\includegraphics[width=\textwidth]{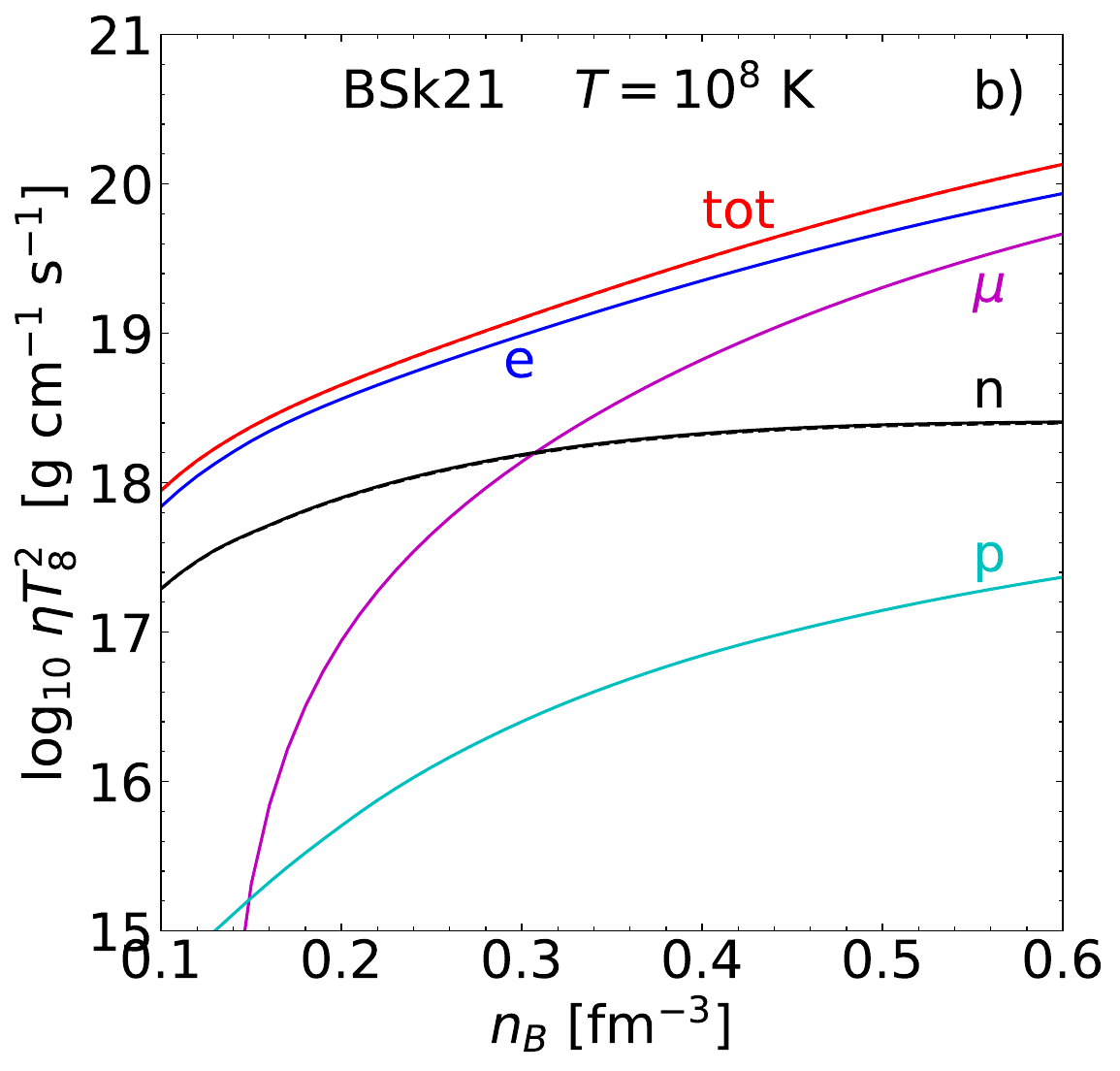}
\end{minipage}
\begin{minipage}{0.32\textwidth}
\includegraphics[width=\textwidth]{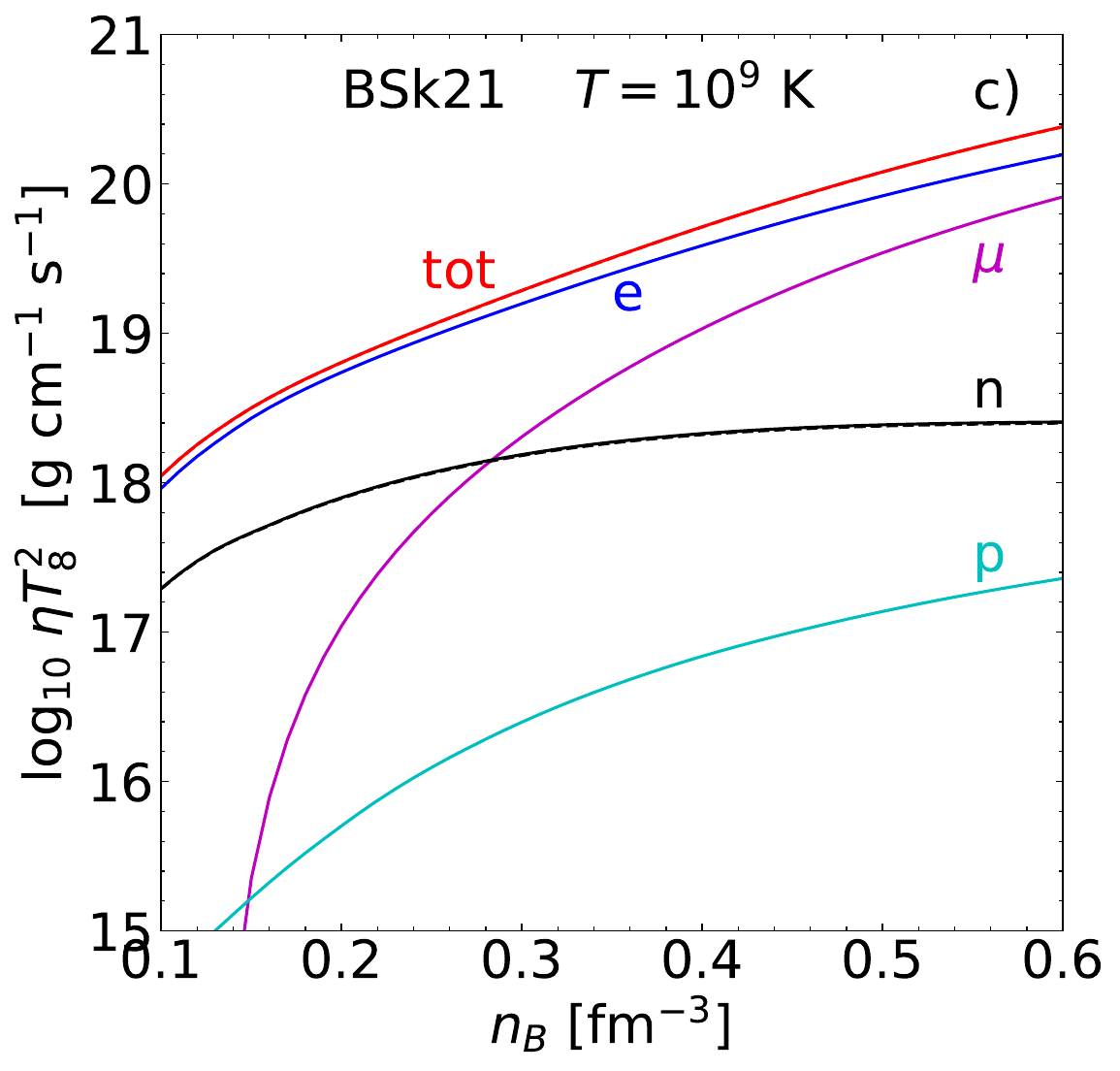}
\end{minipage}
\caption{Partial contributions to the shear viscosity (variational solution) for the beta-stable NS core with the BSk21 equation of state and three values of temperature, $T=10^7$~K (a), $10^8$~K (b), and $10^9$~K (c). The curves are the same as in Figure~\ref{fig:kappa_BSkpart}. However the dashed curves are too close to the solid ones and can not be resolved. }\label{fig:eta_BSkpart}
\end{figure*}

Similar results for the shear viscosity are plotted in Figure~\ref{fig:eta_BSkpart}. Here the combination $\eta T_8^2$ is shown which is temperature-independent in Fermi liquids. Again, in the electromagnetic sector the temperature dependence modifies, $\eta_\ell\propto T^{-5/3}$ in the leading order, so the curves e and $\mu$ are different in the three panels of Figure~\ref{fig:eta_BSkpart}. As opposite to the thermal conductivity case (Figure~\ref{fig:kappa_BSkpart}), leptons dominate the shear viscosity. Only at the lowest densities and temperatures the neutrons give some contribution (Figure~\ref{fig:eta_BSkpart}a). The decoupled solution is shown with the dashed lines which are unresolved in the figure. Therefore this approximation for shear viscosity is even better than for the thermal conductivity. This is especially so since the leptons dominate and the study of the nucleon shear viscosity is more of the academic interest. 

\begin{figure}
    \centering
    \includegraphics[width=\columnwidth]{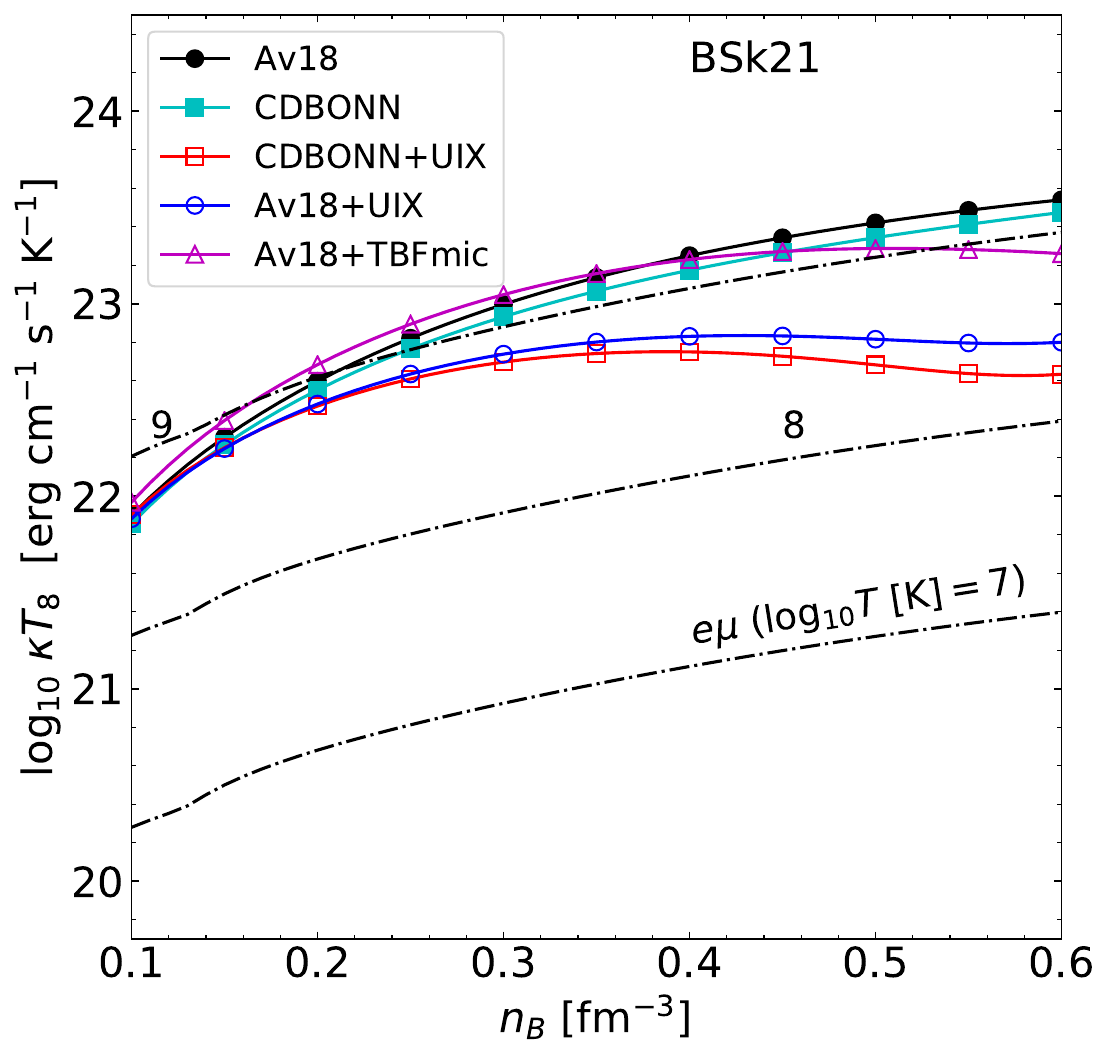}
    \caption{Thermal conductivity for BSk21 EOS. Lines with symbols show interpolated  results for $\kappa_n$ according to practical expressions for different NN interactions (as indicated in the legend). Dash-dotted lines show $\kappa_{e\mu}$ calculated for three values of temperature, whose logarithms are indicated near the curves.}
    \label{fig:kappaBSK}
\end{figure}

According to previous discussion, among the various partial contributions to mean free paths, in practice one only needs  $\lambda_{nn}$ and $\lambda_{np}$. We fitted these quantities by analytical expressions valid in the range $n_B<0.6$~fm$^{-3}$ and $x_p<0.5$. The fits has a form 
\begin{equation}\label{eq:fit}
\frac{\lambda\ T_8^2}{10^{-6}~\mathrm{cm}} =  x_p^\zeta\ \frac{n_B}{n_0}\ 
\sum_{k=0}^3\sum_{m=0}^2 a_{km} \left(\frac{n_B}{n_0}\right)^k x_p^m,
\end{equation}
where $\zeta=0$ for $\lambda^\kappa_{nn}$ and $\lambda^\eta_{nn}$, $\zeta=-1/3$ for $\lambda^\kappa_{np}$, and $\zeta=-1$ for $\lambda^\eta_{np}$ and $\lambda^D_{np}$. \new{Here and in the following fitting formulas $n_0=0.16$~fm$^{-3}$ is adopted.} The polynomial coefficients $a_{km}$ for five interactions considered in this paper  are given in the Table~\ref{tab:fit} in Appendix. Expression~(\ref{eq:fit}) allows one to calculate the effective partial mean free paths for any EOS. The similar fits for the effective masses are provided in Ref.~\cite{Baldo2014PhRvC}. The total neutron mean free path is calculated from the partial mean free paths as
\begin{equation}\label{eq:lambdan_final}
\lambda_n=\left[\lambda_{nn}^{-1}+\lambda_{np}^{-1}\right]^{-1}.
\end{equation}
 
 The neutron thermal conductivity $\kappa_n$ is then calculated using Equation~(\ref{eq:kappa}). We also recommend to multiply Equation~(\ref{eq:kappa}) by a factor  $C_\kappa=1.2$ to correct for the exact solution. The practical expression for the neutron thermal conductivity reads
 \begin{equation}\label{eq:kappa_pract}
    \kappa_n=6.8\times 10^{22}\,T_8 \left(\frac{n_n}{n_0}\right)^{2/3}  \frac{\lambda^\kappa_n}{10^{-6}\ \mathrm{cm}}~\mathrm{erg}~\mathrm{s}^{-1}~\mathrm{cm}^{-1}~\mathrm{K}^{-1}.
\end{equation}
Thermal conductivity for the BSk21 EOS calculated according to Equation~(\ref{eq:kappa_pract}) is shown in Figure~\ref{fig:kappaBSK} for all interactions considered in the paper (different symbols, as indicated in the plot legend). As expected, at low densities $n_B\lesssim 0.3$~fm$^{-3}$ the results for various interactions are quite close. However, at larger densities the curves diverge. With dash-dotted lines we show the lepton contribution to the thermal conductivity for three values of temperature as indicated at the curves. As already anticipated, at $T<10^{8}$~K the neutron contribution dominates for any microscopic model considered. 

\begin{figure}
    \centering
    \includegraphics[width=\columnwidth]{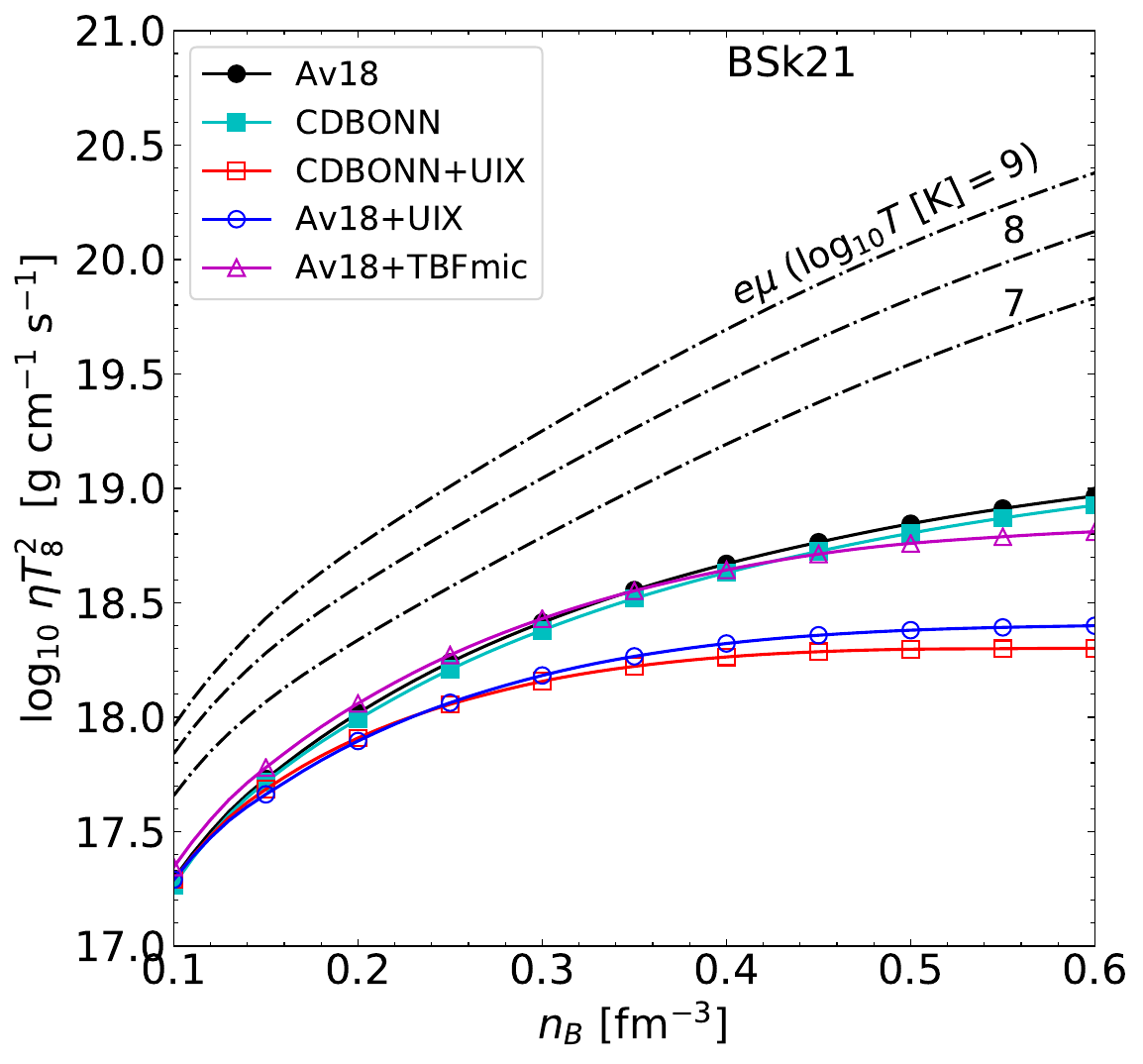}
    \caption{Shear viscosity for BSk21 EOS. Lines with symbols show interpolated  results for $\eta_n$ according to practical expressions for different NN interactions (as indicated in the legend). Dash-dotted lines show $\eta_{e\mu}$ calculated for three values of temperature, whose logarithms are indicated near the curves.}
    \label{fig:etaBSK}
\end{figure}

The neutron shear viscosity in natural units can be written as
\begin{equation}\label{eq:eta_pract}
    \eta_n=5.7\times 10^{17}\, 
    \left(\frac{n_n}{n_0}\right)^{4/3}\, \frac{\lambda^\eta_n}{10^{-6}\ \mathrm{cm}}~\mathrm{g}~\mathrm{s}^{-1}~\mathrm{cm}^{-1}.
\end{equation}
We show the shear viscosity calculated from Equation~(\ref{eq:eta_pract}) in Figure~\ref{fig:etaBSK}. The curves and notations are the same as in Figure~\ref{fig:kappaBSK}. The difference between the different NN interaction models is prominent, although somewhat smaller than in the case of thermal conductivity. Despite the large model uncertainty, $\eta_n$ seems to be negligible in comparison to  $\eta_{e\mu}$ in all cases of practical interest. 

\begin{figure}
    \centering
    \includegraphics[width=\columnwidth]{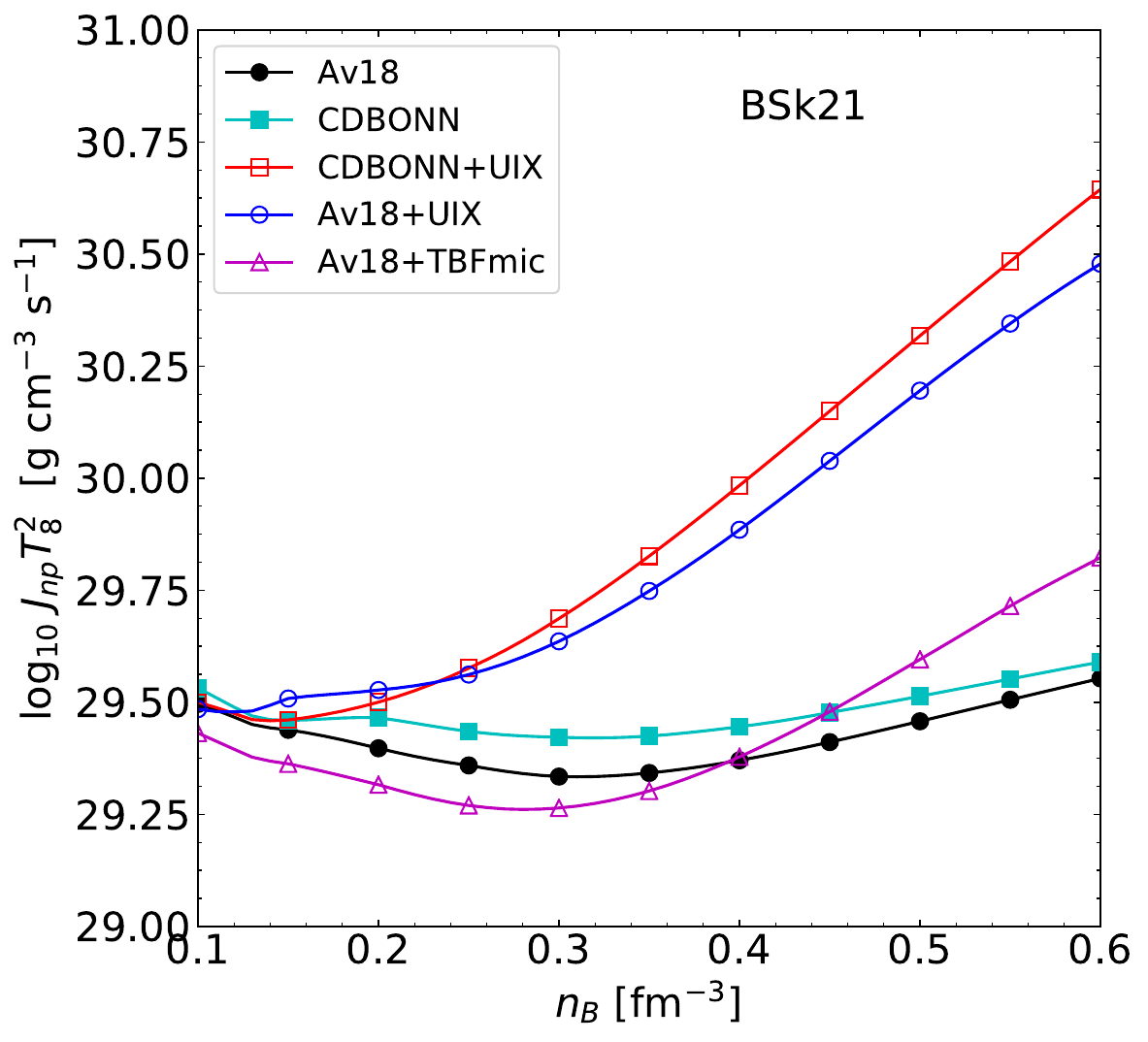}
    \caption{Neutron-proton momentum transfer rate for BSk21 EOS. Lines with symbols show interpolated  results for $J_{np}$ according to practical expressions for different NN interactions (as indicated in the legend).}
    \label{fig:JnpBSK}
\end{figure}

Finally, the similar practical expression for $J_{np}$ (Equation~(\ref{eq:J})) in natural units reads
\begin{equation}\label{eq:Jnp_pract}
J_{np} = 2.8\times 10^{30} \left(\frac{n_n}{n_0}\right)^{4/3} 
\frac{10^{-6}\ \mathrm{cm}}{\lambda^D_{np}}~\mathrm{g}~\mathrm{s}^{-1}~\mathrm{cm}^{-3}.
\end{equation}
In Figure~\ref{fig:JnpBSK}, we show the relaxation rates. Let us stress that this is an important quantity for the magnetic field evolution modelling, since in high magnetic fields it sets the so-called ambipolar diffusion timescale $t_B^{\mathrm{amb}}\sim J_{np} R^2/{\cal B}^2$, where $R$ is the typical scale of the magnetic field and \new{${\cal B}$ is the magnetic field induction}. According to Figure~\ref{fig:JnpBSK}, at large densities the ambipolar diffusion timescale can vary by an order of magnitude depending on the microscopic model used. 
We do not compare here $J_{pn}$ with other (electromagnetic) momentum transfer rates for simplicity. 
Their expressions which include correct plasma screening can be found elsewhere \cite{Shternin2008JETP}.
More detailed discussion can be also found in Ref.~\cite{Dommes2020}.

\section{Conclusions}\label{sec:conclude}
We have analyzed the dependence of the transport coefficients in nucleon cores of the neutron stars on the selection of the nucleon interaction. We employed the non-relativistic Brueckner-Hartree-Fock approach as our microscopic theory. Our results continue the  previous studies in Ref.~\cite{Shternin2013PhRvC} where only the Av18 and Av18+UIX NN interactions were considered.
We provide (Section~\ref{sec:pract}) the  practical  expressions which allow the transport coefficients calculation for any EOS of the nucleon matter, although these expressions are not fully consistent 
since one needs to rely on the specific NN interaction, which can be different from the one on which the EOS is based.
Our general conclusions are as follows
\begin{itemize}
    \item The nucleon contribution to transport coefficients, namely thermal conductivity, shear viscosity, and momentum relaxation rate, strongly depends on the nucleon interaction. At small densities, the difference is modest since all models are restricted by the direct and indirect experimental data.  At large densities, $n_B\sim 0.6$ fm$^{-3}$, the difference between the models considered in this paper can reach an order of magnitude.
    \item The inclusion of the three-body forces in Urbana IX model leads to significant increase in the scattering cross-sections, and, as a result, in decrease in the mean free paths for the quasiparticles and the  transport coefficients. The effect of the microscopic three-body force is less prominent. 
    \item The changes in the effective masses on the Fermi surface are equally important compared with  the changes in the scattering matrix elements in explanation of the obtained difference between the results of calculations for different interactions. Therefore both effects need to be included in practice \cite{Shternin2013PhRvC}.
    \item The main contribution to the nn scattering comes from the ${}^3P_1$  partial wave channel, and to lesser extent from the ${}^3P_2$ and ${}^1S_0$ channels. The difference in the latter channels between the interactions considered in the paper explains most of the observed difference in final results.
    In case of the np scattering, many channels equally contribute, however at high densities ${}^3D_2$ and the coupled ${}^3S_1-{}^3D_1$ partial wave channels are the most important ones.
    \item Despite the differences in the microscopic model, the general conclusion is that the neutron (nucleon) contribution to the thermal conductivity $\kappa$ dominates at $T\gtrsim 10^8$~K, while the lepton contribution to shear viscosity $\eta$ is always dominant. Notice that this conclusion survives also in the case when the proton pairing is taken into account \cite{Shternin2018PRD}.
\end{itemize}

In our study we did not consider the effects of nucleon superfluidity/superconductivity. It is widely accepted that nucleon and protons in a large part of the NS core can be in the paired states, whose critical transition temperatures are uncertain  \cite{LombardoSchulze2001LNP,Haskell2017arXiv,Sedrakian2018arXiv}. Microscopically, the effects of the nucleon pairing on the calculation of transport coefficients are twofold. First,  the quasiparticle spectrum becomes gapped which suppresses the collision probabilities. 
This effect can be incorporated by inclusion of the certain reduction factors \cite{Baiko2001AA, ShterninYakovlev2008}. Second, the transport equation needs to be written for the Bogoluibov quasiparticles, so that the scattering matrix element  which defines the collision integral also modifies (for instance it needs to include the processes related to the nonconservation of such particles),  see, e.g., \cite{Vollhardt1990}. To the best of our knowledge, the modification of the NN scattering matrix elements  by pairing has not been considered  in NS context. Notice that the hydrodynamics equations in the superfluid/superconducting liquid also change and the multifluid picture needs to be invoked. The systematic study of the effects of nuclear pairing on the transport coefficients deserves a separate study.

\acknowledgments
The authors thank Dr. H.-J.~Schulze for providing the subroutine for the TBFmic three-body force calculations. 
A part of this work was done during the PHAROS COST STSM \#MP1304-061014-049724. 
P.S. Thanks the INFN Sez.~di Catania for hospitality. 
The work was partially supported by the Foundation for the Advancement of Theoretical Physics and Mathematics ``BASIS'', grant 17-13-305-1.
\bibliographystyle{apsrev}
\def\apjl{Astrophys. J. Lett.}
\def\apjs{Astroph. J. Suppl. Ser.}
\def\mnras{Mon. Not. R. Astron. Soc.}
\def\aap{Astron. Astrophys.}
\def\plb{Physics Letters {\rm B}}
\def\apss{Astroph. Space Sci.}
\def\pla{Phys. Lett.  A}
\def\ssr{Space Sci. Rev.}
\def\araa{Ann. Rev. Astron. Astrophys.}
\def\aj{Astron. J.}
\def\jphys{J. Phys.}
\def\npa{Nucl. Phys. A}
\def\nphysa{Nucl. Phys. A}
\def\npb{Nucl. Phys.  B}
\def\ijmpe{Int. J. Mod. Phys. E}
\def\ijmpd{Int. J. Mod. Phys. D}
\def\ijmpa{Int. J. Mod. Phys. A}
\bibliography{biblio}

 \begin{widetext}
 \newpage
 \begin{appendix}*
 \section{Parameters of the mean free path fit}
 Here we give the coefficients $a_{km}$ in the fitting expression Equation~(\ref{eq:fit}) for the effective mean free paths. Table~\ref{tab:fit} contains coefficients for calculating $\lambda^\xi_{nb}$, where $\xi = \kappa,\ \eta,\ \mathrm{or}\ D$ and $b=n,\ p$. In all cases, the mean squared relative error of the fit is less than 5\% and the maximal relative fit error does not exceed 15\%.
 
\begin{table}[hb]
 \caption{Parameters of the approximation Equation~(\ref{eq:fit})}\label{tab:fit}
 \begin{tabular}{lc|cccccccccccc}
 \hline\hline
     & $b$& $a_{00}$ &$a_{01}$&$a_{02}$&$a_{10}$&$a_{11}$&$a_{12}$&$a_{20}$&$a_{21}$&$a_{22}$&$a_{30}$&$a_{31}$&$a_{32}$\\
\hline
     Av18\\
     \quad $\kappa$ & n & 
     0.64 & -1.03 & 0.126 & 0.454 & 2.21 & -2.96 & -0.207 & -0.665 & 1.65 & 0.0279 & 0.0612 & -0.225\\
                    & p & 
    0.0588 & -0.0474 & -0.19 & 0.0387 & 0.266 & -0.14 & 0.112 & 0.219 & -0.809 & -0.015 & -0.0738 & 0.192 
        \\  
             \quad$\eta$    & n &   
     2.03 & -3.23 & -0.582 & -0.198 & 7.31 & -4.09 & -0.048 & -3.11 & 2.87 & 0.0189 & 0.401 & -0.454 
      \\
                          & p & 
   0.0467 & -0.114 & -0.072 & -0.0184 & 0.823 & -0.718 & 0.139 & 0.00265 & -0.467 & -0.0231 & -0.0483 & 0.164 
     \\
             \quad  $D$           & p & 
  0.139 & -0.547 & 0.332 & -0.0585 & 2.48 & -3.51 & 0.435 & -0.855 & -0.0522 & -0.0761 & 0.0566 & 0.18 
     \\

   Av18+UIX   \\
   \quad $\kappa$ & n & 
     0.889 & -0.892 & -1.28 & -0.00131 & 3.34 & 0.208 & -0.128 & -1.7 & 0.198 & 0.0223 & 0.236 & -0.0414
     \\
                 & p & 
   -0.00912 & 0.112 & -0.614 & 0.275 & 0.152 & 0.276 & -0.116 & 0.124 & -0.551 & 0.0127 & -0.0369 & 0.124 
        \\  
              \quad $\eta$    & n &   
     2.44 & 0.256 & -6.84 & -1.28 & 3.09 & 7.54 & 0.286 & -1.87 & -2.76 & -0.0229 & 0.275 & 0.333 
      \\
                         & p & 
     -0.0216 & 0.0438 & -0.6 & 0.225 & 0.772 & -0.0735 & -0.0924 & -0.217 & -0.267 & 0.00989 & 0.00768 & 0.0833 
     \\
             \quad  $D$           & p & 
     -0.0192 & -0.577 & 0.454 & 0.593 & 2.73 & -4.84 & -0.233 & -1.07 & 1.7 & 0.0227 & 0.116 & -0.142 
     \\
   Av18 +TBFmic \\
   \quad $\kappa$ & n & 
     0.648 & -1.66 & 0.112 & 0.752 & 5.3 & -4.47 & -0.373 & -2.53 & 2.87 & 0.0446 & 0.328 & -0.428 
     \\
                   & p & 
    0.091 & -1.04 & 1.38 & -0.0549 & 2.94 & -4.47 & 0.244 & -1.41 & 1.72 & -0.0544 & 0.192 & -0.184  
        \\  
             \quad $\eta$    & n &   
     2.31 & -2.84 & -3.73 & -0.417 & 9.2 & 1.28 & -0.0164 & -4.62 & 0.428 & 0.0123 & 0.611 & -0.0884 
      \\
                          & p & 
     0.038 & -0.439 & 0.46 & 0.00716 & 2.1 & -2.83 & 0.157 & -0.671 & 0.7 & -0.0376 & 0.0481 & -0.00242 
     \\
             \quad $D$           & p & 
      0.104 & -1.3 & 1.8 & 0.0604 & 5.33 & -9.15 & 0.455 & -2.35 & 2.87 & -0.111 & 0.283 & -0.189 
     \\
   CDBonn \\
   \quad $\kappa$ & n & 
    0.639 & -1.33 & 0.743 & 0.371 & 2.71 & -4.26 & -0.193 & -0.812 & 2.12 & 0.0262 & 0.0781 & -0.271 
     \\
                   & p & 
    0.01862 & 0.2252 & -0.8008 & 0.1324 & -0.3621 & 1.142 & 0.04245 & 0.2705 & -0.9206 & -0.005907 & -0.05064 & 0.1553 
%
        \\  
             \quad $\eta$    & n &   
          2.0 & -4.51 & 1.87 & -0.201 & 9.63 & -9.67 & -0.0631 & -3.99 & 5.43 & 0.0164 & 0.516 & -0.782 
      \\
                          & p & 
     0.01781 & 0.1886 & -0.8278 & 0.06001 & -4.601$\cdot 10^{-4}$ & 1.054 & 0.07307 & 0.2056 & -0.9559 & -0.01139 & -0.05232 & 0.1836 
     \\
             \quad  $D$           & p & 
     0.1381 & -0.5184 & 0.03195 & 0.003832 & 1.741 & -1.962 & 0.319 & -0.8182 & 0.07881 & -0.05162 & 0.09196 & 0.05515 
     \\
   CDBonn+UIX \\
   \quad $\kappa$ & n & 
    1.24 & 0.0805 & -3.82 & -0.473 & 1.91 & 3.33 & 0.0356 & -0.999 & -1.1 & 0.00373 & 0.132 & 0.134 
     \\
                    & p & 
    -0.0216 & -0.333 & 0.364 & 0.376 & 0.324 & -0.548 & -0.197 & 0.172 & -0.314 & 0.027 & -0.0582 & 0.11 
        \\  
             \quad $\eta$    & n &   
     2.89 & 3.06 & -9.13 & -1.82 & -0.539 & 9.66 & 0.446 & -0.264 & -3.56 & -0.0393 & 0.0545 & 0.44 
      \\
                          & p & 
    -0.0431 & -0.517 & 0.611 & 0.33 & 1.17 & -1.34 & -0.168 & -0.285 & 0.172 & 0.023 & 0.00523 & 0.0383 
     \\
             \quad  $D$           & p & 
      -0.147 & -1.06 & 1.86 & 0.997 & 2.03 & -4.94 & -0.497 & -0.375 & 1.26 & 0.0668 & -0.019 & -0.0294 
     \\
      \end{tabular}
 \end{table}
 
  \end{appendix}
 \end{widetext}

\end{document}